\PassOptionsToPackage{pdfpagelabels=false}{hyperref}
\documentclass[useAMS,usedcolumn,usenatbib]{mnras}

\usepackage{newtxtext,newtxmath}

\usepackage[T1]{fontenc}
\usepackage{ae,aecompl}

\usepackage{graphicx}	
\usepackage{longtable} 
\usepackage{pdflscape} 
\usepackage{footnote}

\newcommand{\isotope}[2]{${}^{#1}$#2}
\newcommand{\msun}{\mbox{$\mathrm{M_{\odot}}$}}
\newcommand{\msunb}{\mbox{$\mathrm{M_{\odot}}$} }

\newcommand{\mwd}{\mbox {$M_{\rm WD}$}}

\newcommand{\gcc}{\mbox {{\rm g~cm$^{-3}$}}}

\newcommand{\cms}{\mbox {{\rm cm~s$^{-1}$}}}
\newcommand{\gccb}{\mbox {{\rm g~cm$^{-3}$}} } 

\newcommand{\cmsb}{\mbox {{\rm cm~s$^{-1}$}} }

\title[Nucleosynthesis in SN Ia]{SNR-calibrated Type Ia supernova models}

\author[Bravo, Badenes \& Mart\'\i nez-Rodr\'\i guez]{
Eduardo Bravo,$^{1}$\thanks{E-mail: eduardo.bravo@upc.edu} 
Carles Badenes,$^{2,3}$ 
H\'ector Mart\'\i nez-Rodr\'\i guez,$^{2}$
\\
$^{1}$E.T.S. Arquitectura del Vall\`es, Universitat Polit\`ecnica de Catalunya, Carrer Pere Serra  
1-15, 08173 Sant Cugat del Vall\`es, Spain\\
$^{2}$Department of Physics and Astronomy and Pittsburgh Particle Physics, Astrophysics and Cosmology Center 
(PITT PACC), \\
University of Pittsburgh, 3941 O'Hara Street, Pittsburgh, PA 15260, USA\\
$^{3}$Institut de Ci\`encies del Cosmos (ICCUB), Universitat de Barcelona (IEEC-UB), Mart\'\i ~Franqu\`es 1, 
E08028 Barcelona, Spain
}

\date{Accepted XXX. Received YYY; in original form ZZZ}

\pubyear{2018}

\begin{document}
\label{firstpage}
\pagerange{\pageref{firstpage}--\pageref{lastpage}}
\maketitle

\begin{abstract}
Current Type Ia supernova (SN Ia) models can reproduce most visible+IR+UV observations. In the X-ray band, the 
determination of elemental abundance ratios in supernova remnants (SNRs) through their spectra has reached 
enough precision to constrain SN Ia models. Mart\'\i nez-Rodr\'\i guez et al have shown 
that the Ca/S mass ratio in SNRs cannot be reproduced with the 
standard nuclear reaction rates for a wide variety of SN Ia models, and suggested that the 
\isotope{12}{C}+\isotope{16}{O} reaction rate could be overestimated by a factor as high as ten. 
We show that the same Ca/S ratio can be 
obtained by simultaneously varying the rates of the reactions \isotope{12}{C}+\isotope{16}{O}, 
\isotope{12}{C}+\isotope{12}{C}, \isotope{16}{O}+\isotope{16}{O}, and 
\isotope{16}{O}$(\gamma,\alpha)$\isotope{12}{C} within the reported uncertainties. We also show that the 
yields of the main products of SN Ia nucleosynthesis do not depend on the details of which rates are 
modified, but can be parametrized by an observational quantity such as Ca/S. 
Using this SNR-calibrated approach, we then proceed to compute a new set of SN Ia models and nucleosynthesis for both Chandrasekhar and 
sub-Chandrasekhar mass progenitors with a one-dimensional hydrodynamics and nucleosynthesis code. 
We discuss the nucleosynthesis of the models as a function of 
progenitor metallicity, mass, and deflagration-to-detonation transition density. 
The yields of each model are almost independent on the reaction rates modified for a common Ca/S ratio.
\end{abstract}


\section{Introduction}\label{s:intro}

Type Ia supernovae (SN Ia) are a fairly homogeneous class of high-luminosity transient phenomena whose 
spectra are characterized by the absence of lines from the most abundant elements in the Cosmos, hydrogen and 
helium, and the presence of a conspicuous silicon line at 6150 \AA\, around maximum light 
\citep{1939min,1941min,1969psk,1973bra,1986whe}. They have been instrumental in the establishment of the 
current cosmological model, characterized by an accelerated expansion of the Universe, the 
measurement of the contribution of dark energy to the energy budget of the Universe, and constraining the equation 
of state of dark energy \citep[e.g.][]{1998rie,1999prl,2004rie,2007wo3,2018sco}. But the imprints of SN Ia go 
beyond their applications as standard candles, and cover aspects as diverse as the evolution of the 
interstellar and intracluster media \citep{1977che,2010saw}, the generation of cosmic rays 
\citep{2005war,2011sin,2015tak,2016car}, or the source of most stable isotopes of the 
elements of the iron-group (IGE) \citep{1987mtt,2009mtt,2017mao,2018mcw,2018pra}. 

The accepted progenitor of SN Ia is a carbon-oxygen (CO) white dwarf (WD) exploding as a consequence of 
destabilization due to mass accretion or whatever other cause. If the WD 
mass is close to the Chandrasekhar limit (Chandra scenario), the explosion can propagate either as a pure 
deflagration \citep{1984nomb} or as a delayed detonation \citep[DDT,][]{1991kho}. If its mass is 
substantially below the Chandrasekhar limit (subCh scenario) the burning front propagates as a pure detonation
\citep{1994woob,2010woo,2010sim}.
Hydrodynamical simulations of WD explosion can explain most of the features in 
the optical light curve and spectra of SN Ia, both in the Chandra and in 
the subCh scenarios \citep{1984nom,1998hoeb,2007woo,2009kas,2013blo,2017blo,2017hoe}. The light curve is 
most sensitive to the kinetic energy imparted to the ejecta and to the mass of \isotope{56}{Ni} synthesized 
in the course of the thermonuclear explosion. The spectra are formed by many lines of intermediate-mass 
elements (IME) and IGE, but the interpretation in terms of the mass of each 
element present in the ejecta is not trivial 
\citep{2005bran,2005sthb,2007ger,2008maz,2011tan,2014sas,2016ash}.

The thermonuclear nature of SN Ia couples the nucleosynthesis with the explosion 
properties, but the direct impact on the optical of all but the major product of these explosions, 
the radioactive isotope \isotope{56}{Ni}, is very limited. As a result, most of the nucleosynthesis 
predictions of current SN Ia models cannot be verified by the available optical data.
Observations of close SN Ia for long periods of time put some constraints on their 
nucleosynthesis, although with large uncertainties \citep{2015dia,2017bot,2017dim,2018gra,2018mag}. The 
X-ray spectra of young SN Ia remnants (SNRs) allow more precision on the determination of mass 
ratios of several IME and IGE \citep{2006bad,2008bad,2013par,2013yan,2015yam,2017dav}. Indeed, \cite{2017mar} have 
determined the mass ratio of calcium to sulfur in several SN Ia remnants with relative errors in the range 
$\sim5\% - 16\%$. Measurements in SNRs are easier because the plasma is optically thin and all the shocked ejecta are 
emitting radiation, allowing for a relatively simple correspondance between the emitted photons and the emitting mass.

Most nucleosynthesis calculations associated with SN Ia models have been performed using the technique 
of post-processing.
In this technique, first a simplified nuclear network is used within the hydrodynamic solver, which allows to obtain the release of nuclear energy with enough accuracy to follow the hydrodynamic evolution of the explosion. In a second step, the detailed nuclear composition of the ejecta is computed using a large nuclear network, i.e. feeding a nuclear kinetic code with the thermodynamical trajectories of different WD 
zones along the explosion, computed using the simplified nuclear network
\citep[e.g.][]{1986thi,2010bra,2016tow,2017leu}. This approach is unavoidable in multi-dimensional simulations of 
SN Ia because of the large requirements on CPU associated with resolving the hydrodynamical part of the 
problem, including flame propagation. However, nowadays it is possible 
to incorporate a large nuclear network in hydrocodes used in one-dimensional 
models of SN Ia \citep{2018mil}.

The aim of the present work is twofold. 
First, we present our one-dimensional SN Ia explosion code, which uses a large nuclear network in the computation of the hydrodynamic evolution, allowing to obtain the detailed nuclear composition directly, without the need of post-processing, and ensuring full coherence between the explosion energetics and the nucleosynthesis.
Second, we address the nucleosynthesis constraints derived from X-ray data on 
SN Ia remnants. Specifically, we explore the ways to match the constraints posed by the observations of 
calcium, argon, and sulfur in several remnants \citep{2017mar}, allowing for reasonable uncertainties in 
several key reaction rates. We identify a combination of rates that satisfies the above constraints, and use 
this SNR-calibration to study the dependence of the nucleosynthesis on parameters such as the 
progenitor metallicity, the WD mass, and the deflagration-to-detonation transition density.

The plan of the paper is as follows. In Section~\ref{s:frame} we explain the general aspects of our method 
and related assumptions. In Section~\ref{s:c+o} we discuss 
the observational constraints posed by X-ray spectra. These constraints allow us to define the 
set of reaction rates adopted in our code. We discuss the nucleosynthesis in 
Section~\ref{s:yields}, and summarize our conclusions in Section~\ref{s:conclu}.  
In Appendix~\ref{appa}, we give the nucleosynthetic yields obtained with the standard set of 
reaction rates and, in 
Appendix~\ref{appb}, we provide further technical details about the characteristics of our code and the way 
in which it incorporates an extensive nuclear network in the hydrodynamics modelling of SN Ia.

\section{Framework}\label{s:frame}

We have computed SN Ia explosion models in spherical symmetry for sub-Chandrasekhar  WD
detonation and Chandra WD DDT. 
Our code integrates 
simultaneously the hydrodynamics, via a PPM solver, and the nuclear network. Here, we describe the
general features of the models, and leave further technical details for appendix~\ref{appb}.

In the subCh scenario, a carbon-oxygen WD with a mass significantly below the Chandrasekhar limit explodes 
through a detonation starting at its center. The detonation may be triggered by an inward shock wave launched 
at the surface by the burning of a tiny layer of helium (not present in our simulations) accreted 
from a degenerate or non-degenerate companion. The detonation might also be the 
consequence of a dynamic event (merging, collision) in a double-degenerate system. The exploding WD is 
characterized by its mass, $M_\mathrm{WD}$, temperature, $T_\mathrm{WD}$, and chemical composition. 
The initial chemical composition is set by the carbon-to-oxygen mass ratio ($\mathrm{C}/\mathrm{O}$) and 
the mass fraction of elements heavier than oxygen. This is equal to the metallicity, $Z$, of the 
progenitor star \citep{2003tim}, because during hydrostatic hydrogen and helium burning the amount of CNO at birth of the progenitor star is transformed into 
$^{22}$Ne. 
Besides \isotope{12}{C}, \isotope{16}{O} and \isotope{22}{Ne}, we complement the initial chemical composition with metals with baryon number $23\le A\le100$, which we assume are present in solar proportions \citep{1993grv} with respect to $Z$. 

In the Chandra DDT scenario, a carbon-oxygen WD with a mass close to the Chandrasekhar limit explodes, 
starting by a subsonic flame (deflagration) at or near its center. After consumption of a small fraction of 
the WD mass and expansion by a factor of $\sim2-6$ in radius, the burning fronts turns into a detonation that 
processes most of the star. The origin of the deflagration may be accretion of matter from a non-degenerate 
companion or from the debris of a degenerate companion after a merging event. Either way, the 
explosion is preceded by a long ($\sim10^3$~yrs) phase of slow carbon burning (usually referred to as simmering or smoldering), in which the 
chemical composition is altered with respect to that of the WD at its birth. These models
are characterized by the value of the density ahead the flame at the moment in which the deflagration-to-detonation transition is 
induced, $\rho_\mathrm{DDT}$, and also by the initial central density, $\rho_\mathrm{c}$, temperature, $T_\mathrm{WD}$, 
and chemical composition, which we specify as in the subCh models. However, due to the chemical processing 
during carbon simmering, the actual composition of matter is expected to differ from solar 
proportions. The amount of carbon consumed during simmering may be as large as 
0.026--0.036~\msunb \citep{2017pie}, depending on the initial metallicity, but its precise value is 
affected by the numerical treatment of the URCA process \citep[Piersanti, priv. comm.;][]{2016mar,2017sch}, so we do not include this 
effect in our initial models. We warn that the actual metallicity of the main-sequence 
progenitor of the exploding WD in the Chandra DDT scenario may be different from the $Z$ of the 
models as reported in Table~\ref{tab1}.

In all our models, we adopt $\mathrm{C}/\mathrm{O}=1$\footnote{In this paper, we adopt the convention that 
the ratio of two element symbols, e.g. C/O or Ca/S, makes reference to the ratio of the corresponding masses throughout the ejecta.}
and $T_\mathrm{WD}=10^8$~K. All the initial models are 
built in hydrostatic equilibrium using the specified chemical composition. The central density of our Chandra 
DDT models is $\rho_\mathrm{c}=3\times10^9$~\gcc, as suggested by models of the carbon simmering phase \citep{2016mar}.

The thermonuclear reaction rates used in the simulations are those reccommended by the REACLIB compilation 
\citep{2010cyb}, with electron screening in strong, intermediate and weak regimes, while weak interaction 
rates are adopted from \cite{1982ful,1994oda,2000gmp,2003pru}. The exception is the 
\isotope{12}{C}+\isotope{16}{O} reaction.\footnote{See Section~\ref{s:losotros} for an exploration of other modifications of reaction rates.}
We computed models with either 
the 'standard' \isotope{12}{C}+\isotope{16}{O} reaction rate \citep[][hereafter CF88]{1988cau}, or the same rate scaled down by a factor 
$\left(1-\xi_\mathrm{CO}\right)$, with $\xi_\mathrm{CO}=0.9$ to reproduce the Ca/S mass ratio 
in SNRs \citep[][see also Section~\ref{s:c+o} for more details]{2017mar}. In both cases, we have adopted the 
CF88 branching ratios for the neutron, 
proton, and $\alpha$ output channels.

\begin{figure}
   \includegraphics[width=\columnwidth]{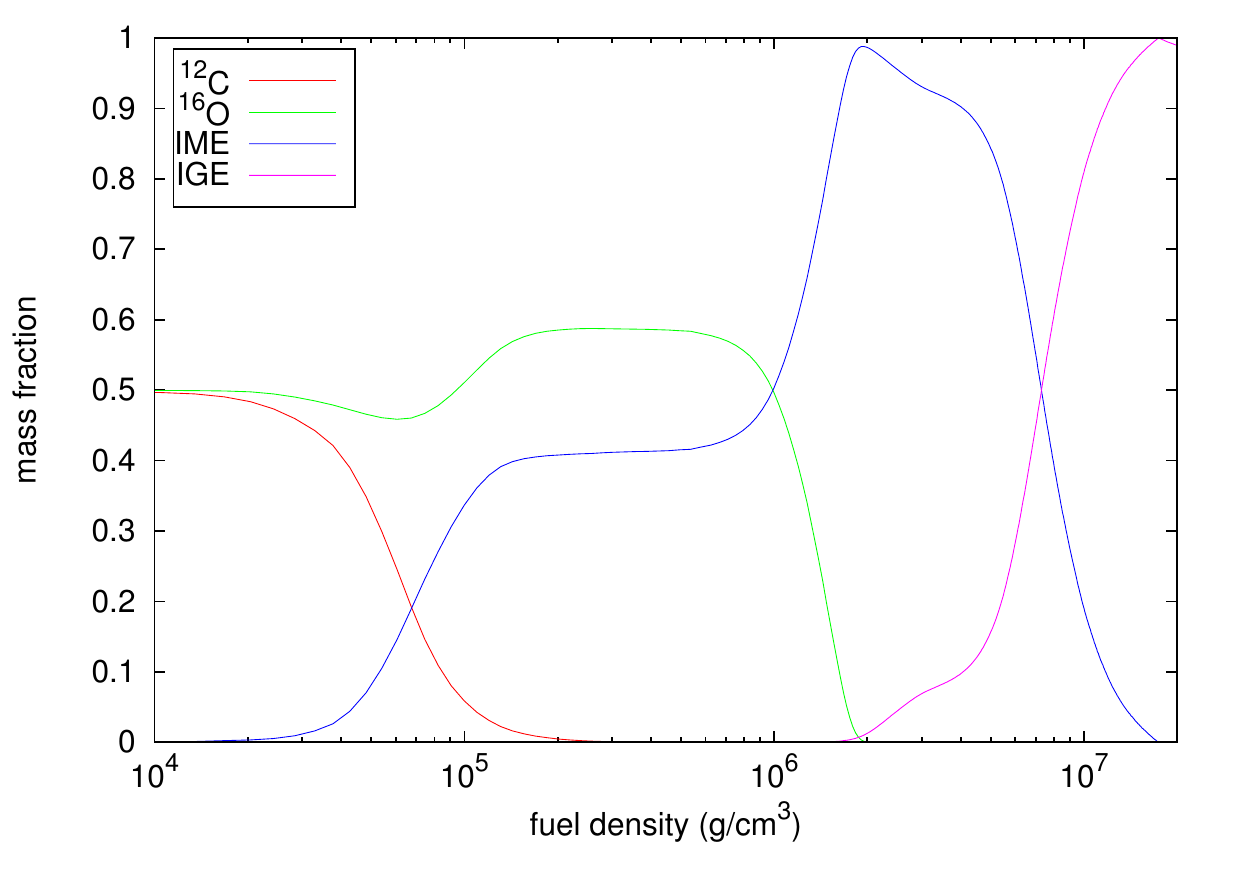}
    \caption{
    Nucleosynthetic yields as a function of density resulting from detonation of C/O in model 1p06\_Z2p25e-3\_std. 
    We show the main constituents of the ejected matter, carbon, oxygen, IME, and IGE. Nevertheless, for fuel densities in the range $4\times10^4 - 10^5$~\gcc, the abundance of neon (not plotted here) may be as large as $\sim16$\% by mass.
    }
    \label{fig1}
\end{figure}

The propagation model for the burning front is different depending on whether it is subsonic or supersonic. 
In the first case, i.e for deflagration waves, the front is propagated at a fixed fraction of the local 
sound speed, $v_\mathrm{def}=0.03v_\mathrm{sound}$. Once a detonation has been initiated, the front velocity is not prescribed,  
and the detonation advances as a result of the associated 
shock wave and the heat released by nuclear burning. The resulting detonation 
velocity is close to the Chapman-Jouguet value for the densities of interest,
i.e. $v_\mathrm{det}\sim\left(1.1-1.3\right)\times10^9$~\cmsb for fuel densities 
$\rho_\mathrm{fuel}\le4\times10^7$~\gcc \citep[][]{1999gam}. The nucleosynthesis  
is mainly determined by the fuel density, see Fig.~\ref{fig1}. In this figure, we have 
grouped all IME, here defined as those elements between and including magnesium and scandium, and all IGE to facilitate comparison with existing 
literature, e.g. figure A.1 in \cite{2010fin}, with which the agreement is quite satisfactory.

We present the hydrodynamic and nucleosynthesis output from 100 models, obtained by 
combining 
five values of $\rho_\mathrm{DDT}$, five values of $Z$, and two parametrizations of the 
\isotope{12}{C}+\isotope{16}{O} reaction rate, for the Chandra models, and five values of $M_\mathrm{WD}$, 
together with the same five values of $Z$ and the two parametrizations of the \isotope{12}{C}+\isotope{16}{O} reaction rate, 
for the subCh models. 
Chandra models are named starting by 'ddt', then the value of $\rho_\mathrm{DDT}$ in units of $10^7$~\gcc, then '\_Z' followed by the progenitor metallicity, then either '\_$\xi_\mathrm{CO}$0p9' for the models ran with $\xi_\mathrm{CO}=0.9$ or '\_std' for the models ran with the standard CF88 \isotope{12}{C}+\isotope{16}{O} reaction rate. For instance, model 'ddt1p2\_Z2p25e-4\_$\xi_\mathrm{CO}$0p9' belongs to the delayed-detonation of a Chandrasekhar-mass WD with $\rho_\mathrm{DDT}=1.2\times10^7$~\gcc, metallicity $Z=2.25\times10^{-4}$, and $\xi_\mathrm{CO}=0.9$. On the other hand, subCh models are named starting with the value of the WD mass, in solar masses, then the metallicity and the treatment of the \isotope{12}{C}+\isotope{16}{O} reaction rate, as in Chandra models. For instance, model '1p06\_Z2p25e-3\_std' belongs to the detonation of a sub-Chandrasekhar mass WD with $M_\mathrm{WD}=1.06$~\msun, metallicity $Z=2.25\times10^{-3}$, and $\xi_\mathrm{CO}=0$, i.e. the standard CF88 \isotope{12}{C}+\isotope{16}{O} reaction rate.

The complete list of models is given in Table~\ref{tab1}, as well as the final kinetic 
energy, $K$, and the ejected mass of \isotope{56}{Ni}. We emphasize that the central density reported for the 
subCh models is a result of the construction of the initial models in hydrostatic equilibrium for given 
$M_\mathrm{WD}$ and $Z$. As a result, it reflects a slight dependence on the WD metallicity, especially for 
the most neutronized progenitors. 

\begin{table*}
 \caption{Characteristics of the computed SN Ia models.}
 \label{tab1}
\renewcommand{\thefootnote}{\alph{footnote}}
 \begin{tabular}{@{}lllllll@{}}
 \hline\hline
 \multicolumn{3}{l}{\rule{0pt}{3ex}Chandrasekhar-mass DDT} & \multicolumn{2}{c}{\hrulefill models with 
$\xi_\mathrm{CO}=0.9$\hrulefill} & 
\multicolumn{2}{l}{\hrulefill models with $\xi_\mathrm{CO}=0$\hrulefill} \\
 $\rho_\mathrm{DDT}$ & $Z$ & $\rho_c$ & $K$ & $M ($\isotope{56}{Ni}$)$ & $K$ & $M 
($\isotope{56}{Ni}$)$ \\
 (\gcc) & & (\gcc) & ($10^{51}$~erg) & (\msun) & ($10^{51}$~erg) & (\msun) \\
 \hline
 $1.2\times10^7$  & $2.25\times10^{-4}$ & $3.0\times10^9$   & 1.193 & 0.316 & 
 1.170 & 0.269 \\
 $1.2\times10^7$  & $2.25\times10^{-3}$  & $3.0\times10^9$   & 1.185 & 0.303 & 
 1.169 & 0.266 \\
 $1.2\times10^7$  & $9.00\times10^{-3}$    & $3.0\times10^9$   & 1.182 & 0.293 & 
 1.165 & 0.251 \\
 $1.2\times10^7$  & $2.25\times10^{-2}$   & $3.0\times10^9$   & 1.158 & 0.250 & 
 1.139 & 0.214 \\
 $1.2\times10^7$  & $6.75\times10^{-2}$   & $3.0\times10^9$   & 1.138 & 0.188 & 
 1.131 & 0.185 \\
 $1.6\times10^7$  & $2.25\times10^{-4}$ & $3.0\times10^9$   & 1.328 & 0.517 & 
 1.317 & 0.491 \\
 $1.6\times10^7$  & $2.25\times10^{-3}$  & $3.0\times10^9$   & 1.323 & 0.510 & 
 1.314 & 0.487 \\
 $1.6\times10^7$  & $9.00\times10^{-3}$    & $3.0\times10^9$   & 1.307 & 0.476 & 
 1.292 & 0.443 \\
 $1.6\times10^7$  & $2.25\times10^{-2}$   & $3.0\times10^9$   & 1.287 & 0.434 & 
 1.285 & 0.426 \\
 $1.6\times10^7$  & $6.75\times10^{-2}$   & $3.0\times10^9$   & 1.277 & 0.366 & 
 1.277 & 0.368 \\
 $2.4\times10^7$  & $2.25\times10^{-4}$ & $3.0\times10^9$   & 1.447 & 0.750 & 
 1.433 & 0.716 \\
 $2.4\times10^7$  & $2.25\times10^{-3}$  & $3.0\times10^9$   & 1.443 & 0.743 & 
 1.429 & 0.710 \\
 $2.4\times10^7$  & $9.00\times10^{-3}$    & $3.0\times10^9$   & 1.429 & 0.704 & 
 1.420 & 0.685 \\
 $2.4\times10^7$  & $2.25\times10^{-2}$   & $3.0\times10^9$   & 1.413 & 0.663 & 
 1.409 & 0.655 \\
 $2.4\times10^7$  & $6.75\times10^{-2}$   & $3.0\times10^9$   & 1.389 & 0.549 & 
 1.385 & 0.546 \\
 $2.8\times10^7$  & $2.25\times10^{-4}$ & $3.0\times10^9$   & 1.468 & 0.804 & 
 1.457 & 0.780 \\
 $2.8\times10^7$  & $2.25\times10^{-3}$  & $3.0\times10^9$   & 1.463 & 0.794 & 
 1.451 & 0.767 \\
 $2.8\times10^7$  & $9.00\times10^{-3}$    & $3.0\times10^9$   & 1.453 & 0.765 & 
 1.447 & 0.754 \\
 $2.8\times10^7$  & $2.25\times10^{-2}$   & $3.0\times10^9$   & 1.437 & 0.721 & 
 1.431 & 0.707 \\
 $2.8\times10^7$  & $6.75\times10^{-2}$   & $3.0\times10^9$   & 1.412 & 0.595 & 
 1.410 & 0.596 \\
 $4.0\times10^7$  & $2.25\times10^{-4}$ & $3.0\times10^9$   & 1.507 & 0.909 & 
 1.500 & 0.896 \\
 $4.0\times10^7$  & $2.25\times10^{-3}$  & $3.0\times10^9$   & 1.503 & 0.902 & 
 1.497 & 0.891 \\
 $4.0\times10^7$  & $9.00\times10^{-3}$    & $3.0\times10^9$   & 1.493 & 0.872 & 
 1.487 & 0.859 \\
 $4.0\times10^7$  & $2.25\times10^{-2}$   & $3.0\times10^9$   & 1.478 & 0.824 & 
 1.475 & 0.817 \\
 $4.0\times10^7$  & $6.75\times10^{-2}$   & $3.0\times10^9$   & 1.456 & 0.689 & 
 1.454 & 0.688 \\
 \hline
 \multicolumn{3}{l}{\rule{0pt}{3ex}sub-Chandrasekhar detonation} & \multicolumn{2}{c}{\hrulefill models with 
$\xi_\mathrm{CO}=0.9$\hrulefill} & 
\multicolumn{2}{l}{\hrulefill models with $\xi_\mathrm{CO}=0$\hrulefill} \\
 \mwd & $Z$ & $\rho_c$ & $K$ & $M ($\isotope{56}{Ni}$)$ & $K$ & $M 
($\isotope{56}{Ni}$)$ \\
 (\msun) & & (\gcc) & ($10^{51}$~erg) & (\msun) & ($10^{51}$~erg) & (\msun) \\
 \hline
 0.88 & $2.25\times10^{-4}$ & $0.15\times10^8$  & 0.926 & 0.191 & 
 0.907 & 0.150 \\
 0.88 & $2.25\times10^{-3}$  & $0.15\times10^8$  & 0.921 & 0.182 & 
 0.905 & 0.145 \\
 0.88 & $9.00\times10^{-3}$    & $0.15\times10^8$  & 0.917 & 0.169 & 
0.904 & 0.138 \\
 0.88 & $2.25\times10^{-2}$   & $0.15\times10^8$  & 0.913 & 0.155 &  
0.902 & 0.133 \\
 0.88 & $6.75\times10^{-2}$   & $0.16\times10^8$  & 0.926 & 0.139 &  
0.920 & 0.133 \\
 0.97 & $2.25\times10^{-4}$ & $0.26\times10^8$  & 1.160 & 0.457 &  
1.140 & 0.427 \\
 0.97 & $2.25\times10^{-3}$  & $0.26\times10^8$  & 1.150 & 0.449 &  
1.140 & 0.421 \\
 0.97 & $9.00\times10^{-3}$    & $0.26\times10^8$  & 1.150 & 0.433 & 
1.140 & 0.410 \\
 0.97 & $2.25\times10^{-2}$   & $0.27\times10^8$  & 1.140 & 0.412 &  
1.130 & 0.395 \\
 0.97 & $6.75\times10^{-2}$   & $0.28\times10^8$  & 1.150 & 0.363 &  
1.140 & 0.358 \\
 1.06 & $2.25\times10^{-4}$ & $0.47\times10^8$  & 1.330 & 0.706 &  
1.320 & 0.685 \\
 1.06 & $2.25\times10^{-3}$  & $0.47\times10^8$  & 1.330 & 0.699 &  
1.320 & 0.679 \\
 1.06 & $9.00\times10^{-3}$    & $0.48\times10^8$  & 1.320 & 0.680 & 
1.320 & 0.664 \\
 1.06 & $2.25\times10^{-2}$   & $0.49\times10^8$  & 1.320 & 0.650 &  
1.310 & 0.638 \\
 1.06 & $6.75\times10^{-2}$   & $0.52\times10^8$  & 1.320 & 0.569 &  
1.320 & 0.565 \\
 1.10 & $2.25\times10^{-4}$ & $0.63\times10^8$  & 1.400 & 0.807 &  
1.390 & 0.791 \\
 1.10 & $2.25\times10^{-3}$  & $0.63\times10^8$  & 1.390 & 0.801 &  
1.390 & 0.785 \\
 1.10 & $9.00\times10^{-3}$    & $0.64\times10^8$  & 1.390 & 0.781 & 
1.380 & 0.769 \\
 1.10 & $2.25\times10^{-2}$   & $0.65\times10^8$  & 1.380 & 0.748 &  
1.380 & 0.739 \\
 1.10 & $6.75\times10^{-2}$   & $0.71\times10^8$  & 1.380 & 0.653 &  
1.380 & 0.650 \\
 1.15 & $2.25\times10^{-4}$ & $0.94\times10^8$  & 1.470 & 0.928 &  
1.460 & 0.918 \\
 1.15 & $2.25\times10^{-3}$  & $0.94\times10^8$  & 1.460 & 0.922 &  
1.460 & 0.913 \\
 1.15 & $9.00\times10^{-3}$    & $0.95\times10^8$  & 1.460 & 0.901 & 
1.460 & 0.894 \\
 1.15 & $2.25\times10^{-2}$   & $0.98\times10^8$  & 1.450 & 0.865 &  
1.450 & 0.859 \\
 1.15 & $6.75\times10^{-2}$   & $1.07\times10^8$  & 1.450 & 0.753 &  
1.450 & 0.751 \\
 \hline\hline
 \end{tabular}
\end{table*}

\begin{figure}
   \includegraphics[width=\columnwidth]{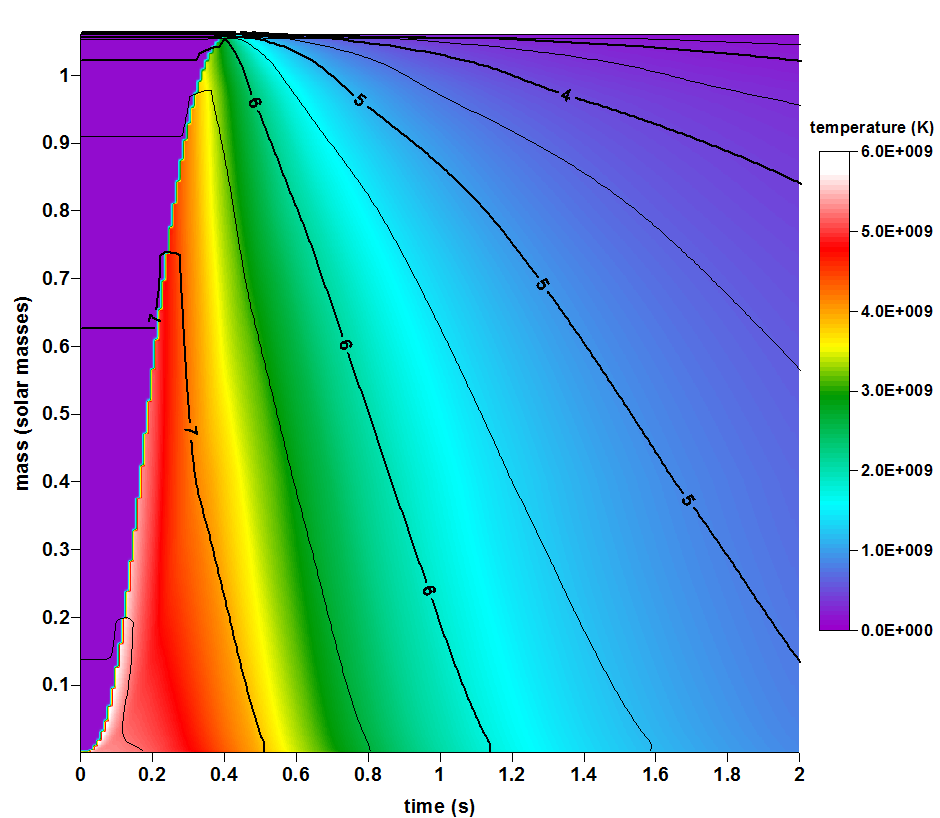}
    \caption{Evolution of temperature (color) and density (contours) as a function of time and mass 
coordinate in a sub-Chandrasekhar detonation model with $M_\mathrm{WD}=1.06$~\msunb and metallicity 
$Z=0.009$, model 1p06\_Z9e-3\_std. The contour labels give the logarithm of the density.
    \label{fig2}}
\end{figure}

\begin{figure}
   \includegraphics[width=\columnwidth]{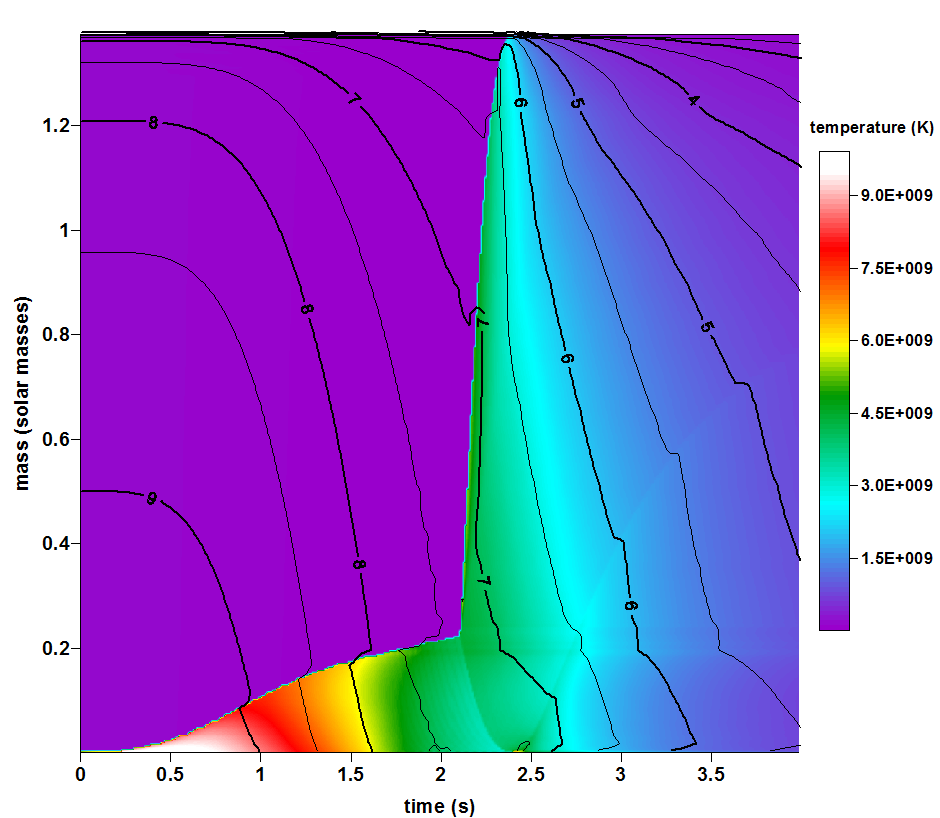}
    \caption{Same as Fig.~\ref{fig2}, but for a Chandrasekhar-mass delayed detonation model 
with deflagration-to-detonation transition density 
$\rho_\mathrm{DDT}=2.4\times10^7$~\gccb and metallicity $Z=0.009$, model ddt2p4\_Z9e-3\_std. Note that the 
color scale, as well as the x and y axes, are different than those of the previous figure. 
    \label{fig3}}
\end{figure}

In Figs.~\ref{fig2} and \ref{fig3}, we show the evolution of a sample of models, one subCh and one Chandra. 
Density (contours) and temperature (color) are shown as functions of time and mass 
coordinate. The pure detonation nature of the subCh models make their evolution relatively simple: matter 
burns at the density it has in the initial model, as shown by the horizontal density contours before 
arrival of the detonation. The mass 
consumption rate is: $\dot{M}=4\pi\rho r^2 v_\mathrm{det}$, which scales as $r^2$ near the center (constant 
fuel density) and declines after the front reaches regions with smaller density, hence smaller 
$v_\mathrm{det}$. The detonation wave approaches the outermost layers of the WD after $\sim0.5$~s. The 
maximum temperature attained behind the detonation front is a declining function of fuel density, going from 
$\sim6\times10^9$~K close to the center to $<3\times10^9$~K close to the surface. 
Models characterized by a smaller central density also reach lower maximum temperatures. 

The evolution of the Chandra DDT model is more involved. The explosion starts at the center of 
the WD as a deflagration and propagates for two seconds, consuming 0.22~\msunb before changing into a 
detonation. Since the deflagration is subsonic, the star has time to expand before the arrival of the burning 
front. The expansion is communicated gradually from the innermost to the outermost layers (density contours start bending downwards), and becomes 
global $\sim1$~s after thermal runaway. Once initiated, the detonation reaches almost the surface of 
the WD without apparent interference from the hydrodynamic processes inside. However, a close inspection of 
Fig.~\ref{fig3} reveals some ripples in the density contours accompanied by subtle changes in 
temperature. The last are due to an inwards-moving shock wave, launched from the location of the detonation 
initiation, that reaches the center of the WD at $\sim2.4$~s and rebounds, moving outwards thereafter. Since 
our code solves simultaneously hydrodynamics and nuclear network, it incorporates the effects of 
all these waves on the evolution of the explosion.

\section{Calcium, sulfur and the fusion of carbon and oxygen}\label{s:c+o}

\begin{figure}
   \includegraphics[width=\columnwidth]{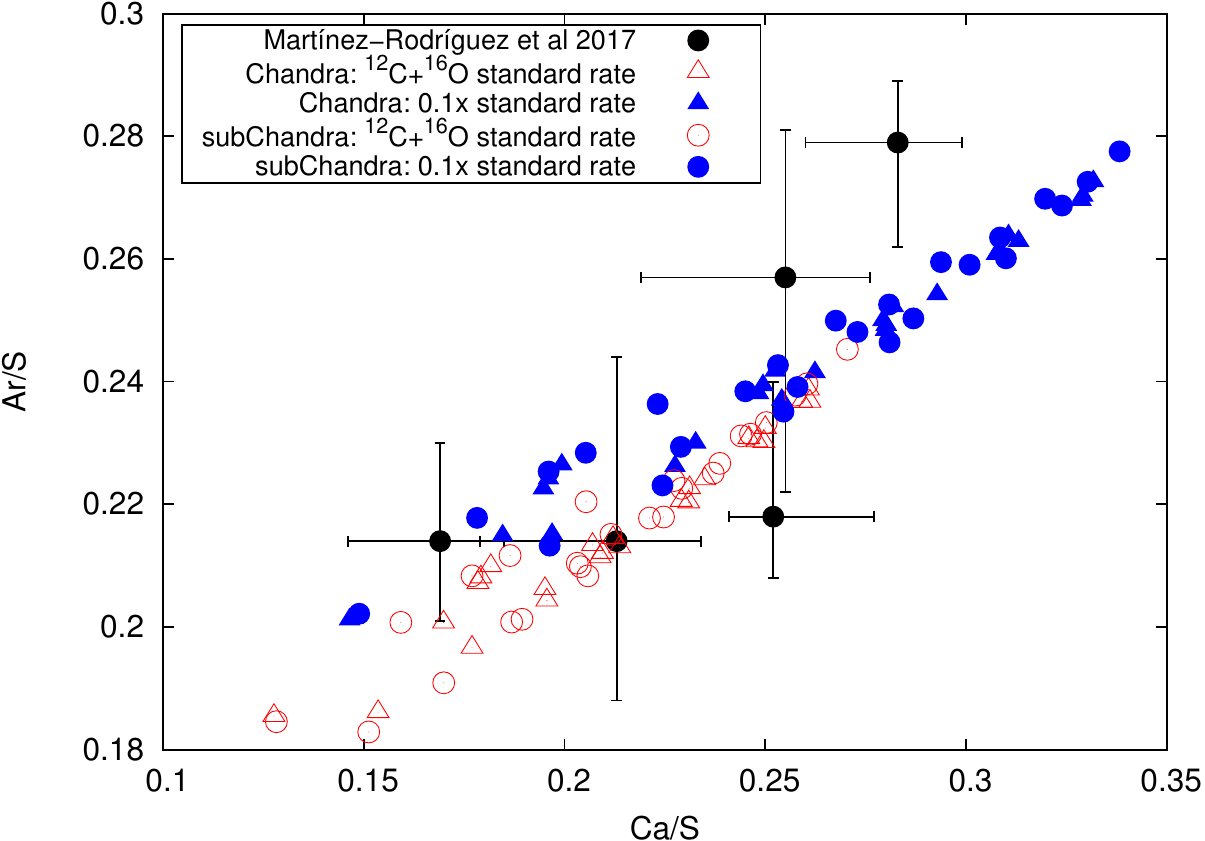}
    \caption{Measured abundance ratio of argon to sulfur vs calcium to sulfur in supernova remnants (big black 
dots with errorbars) compared with the predictions of our model set using either the standard 
\isotope{12}{C}+\isotope{16}{O} reaction rate (empty magenta symbols) or the same rate scaled down by a factor 10, 
i.e. $\xi_\mathrm{CO}=0.9$ (solid blue symbols). 
The uppermost black point belongs to the {\sl Kepler} SNR.
    \label{fig4}}
\end{figure}

\cite{2017mar} argued that the Ca/S ratios measured in the X-ray spectra of Type Ia SNRs span values that 
are not reproduced by most SN Ia explosion models. These authors  
speculate that the cause may be the overestimation of the  
\isotope{12}{C}+\isotope{16}{O} reaction rate by a factor $\sim10$, i.e. $\xi_\mathrm{CO}=0.9$. Fig.~\ref{fig4} 
shows Ca/S and Ar/S obtained with our SN Ia models, with either the standard rate of this reaction or 
$\xi_\mathrm{CO}=0.9$, compared with the abundance ratios measured by
\cite{2017mar}. 
Models calculated with the standard rate of the 
\isotope{12}{C}+\isotope{16}{O} reaction are unable to reproduce both the Ar/S and the Ca/S ratios, in particular that of the {\sl Kepler} SNR. 
On the other hand, 
models with $\xi_\mathrm{CO}=0.9$ cover the whole range of observational points. However, the reduction of 
the rate of the reaction \isotope{12}{C}+\isotope{16}{O} by a factor of ten may be beyond the current 
experimental uncertainty of this rate, which \cite{2018she} estimate on the order of $50$\% at the 
temperatures of interest for explosive oxygen burning, $T\sim(3.5$--$5)\times10^9$~K.

\subsection{Impact of the \isotope{12}{C}+\isotope{16}{O} reaction rate}\label{s:impact}

\begin{figure}
   \includegraphics[width=\columnwidth]{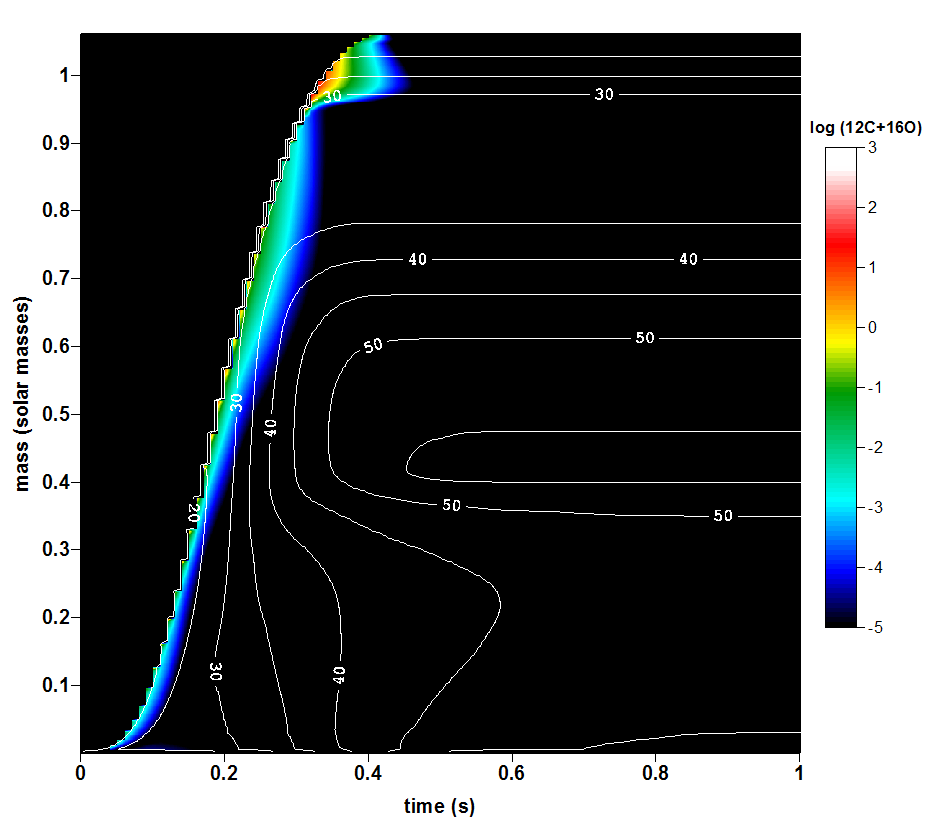}
    \caption{Time evolution of the mean molecular weight (contours) and the mass flux of the 
\isotope{12}{C}+\isotope{16}{O} reaction, given by
$28\mathrm{N}_\mathrm{Av}\rho <\sigma\cdot v> Y_\mathrm{^{12}C} Y_\mathrm{^{16}O}$
(color, logarithmic scale), for model 1p06\_Z9e-3\_std.
    \label{fig6}}
\end{figure}

Scaling-down the \isotope{12}{C}+\isotope{16}{O} reaction rate with $\xi_\mathrm{CO}=0.9$ affects unevenly different elements present in 
the ejecta of SN Ia. The impact of this reaction rate on the yield of a given element
can be understood by comparing the locations where the element is produced with the places where the reaction 
is most active. Fig.~\ref{fig6} shows the mass flux associated with the 
\isotope{12}{C}+\isotope{16}{O} reaction as a function of mass coordinate and time, for model 1p06\_Z9e-3\_std. The flux is large in two 
regions, one between mass coordinates $\sim0.6$ and $\sim0.85$~\msun, and the other outside of 
$\sim0.97$~\msun, where the final mean molecular weights are $35$--$50$ and $<30$, respectively. 
Thus, we expect that IMEs and lighter elements are most sensitive to the \isotope{12}{C}+\isotope{16}{O} reaction rate.

\begin{figure*}
   \includegraphics[width=\textwidth]{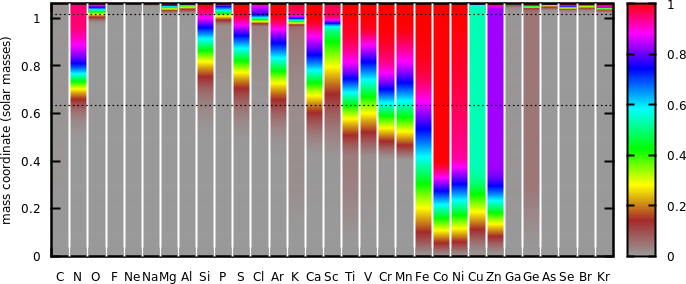}
    \caption{Final distribution of each element through the ejecta in model 1p06\_Z9e-3\_std. The color 
represents the cumulated mass of each element, starting from the center of the star, normalized to the total 
ejected mass of the same element. All elements go through light brown at the center to red at the surface, 
although this is not apparent for several elements whose yield is strongly concentrated in the outermost 
layers of the ejecta, e.g. carbon or gallium. The two black dashed lines enclose the region where the 
maximum temperature achieved by matter is in the range $3.5\le T_\mathrm{9,max}\le 5.0$, in units of $10^9$~K.
    \label{fig7}}
\end{figure*}

Fig.~\ref{fig7} is a map of the final distribution of each element in the ejecta of model 1p06\_Z9e-3\_std, 
normalized to the total yield of the same element, thus, going from 0 at the 
center (brown) to 1 at the surface (red). The two horizontal dotted lines limit the region where conditions
for explosive oxygen burning are achieved. 
It can be seen that several groups of elements are synthesized in similar regions of the white dwarf. For instance, most odd-Z IMEs are produced in the outermost 10-15\% of the star, while most even-Z IMEs and the lightest IGEs, from titanium to manganese, are created in the outer half of the white dwarf.

Elements from the iron-group are mostly 
produced in the innermost regions of the WD, and their yield is hardly influenced by the rate of the 
\isotope{12}{C}+\isotope{16}{O} reaction. On the other hand, nitrogen, silicon, argon, calcium, and 
scandium are produced in the zones of maximal influence of this reaction, and within the regions that experience explosive
oxygen burning. 
Sulfur is particular, in the sense that its region of maximal productivity coincides with the gap in the mass flux associated with the 
\isotope{12}{C}+\isotope{16}{O} reaction (Fig.~\ref{fig6}), i.e. more than half the yield of sulfur in Fig.~\ref{fig7} is concentrated in between the mass range from 0.85~\msunb to 0.97~\msun.
The yields of other elements that 
are preferentially synthesized in the outermost regions of the WD are sensitive to this reaction rate, but are not 
a product of explosive oxygen burning and
their abundances are, in general, very small. 
This is the case, for instance, of potassium and chlorine.

\begin{figure*}
   \includegraphics[width=\textwidth]{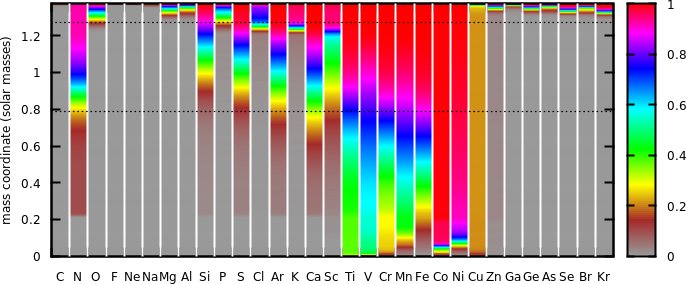}
    \caption{Same as Fig.~\ref{fig7} but for model ddt2p4\_Z9e-3\_std.
    \label{fig8}}
\end{figure*}

Figure~\ref{fig8} shows the same kind of map, pertaining to a Chandra model. Qualitatively, the distribution 
of yields is similar to that of the subCh model, although in the Chandra model the lightest IGE are 
synthesized closer to the center, while copper and zinc are synthesized very close to the 
surface of the WD. 

\begin{figure}
   \includegraphics[width=\columnwidth]{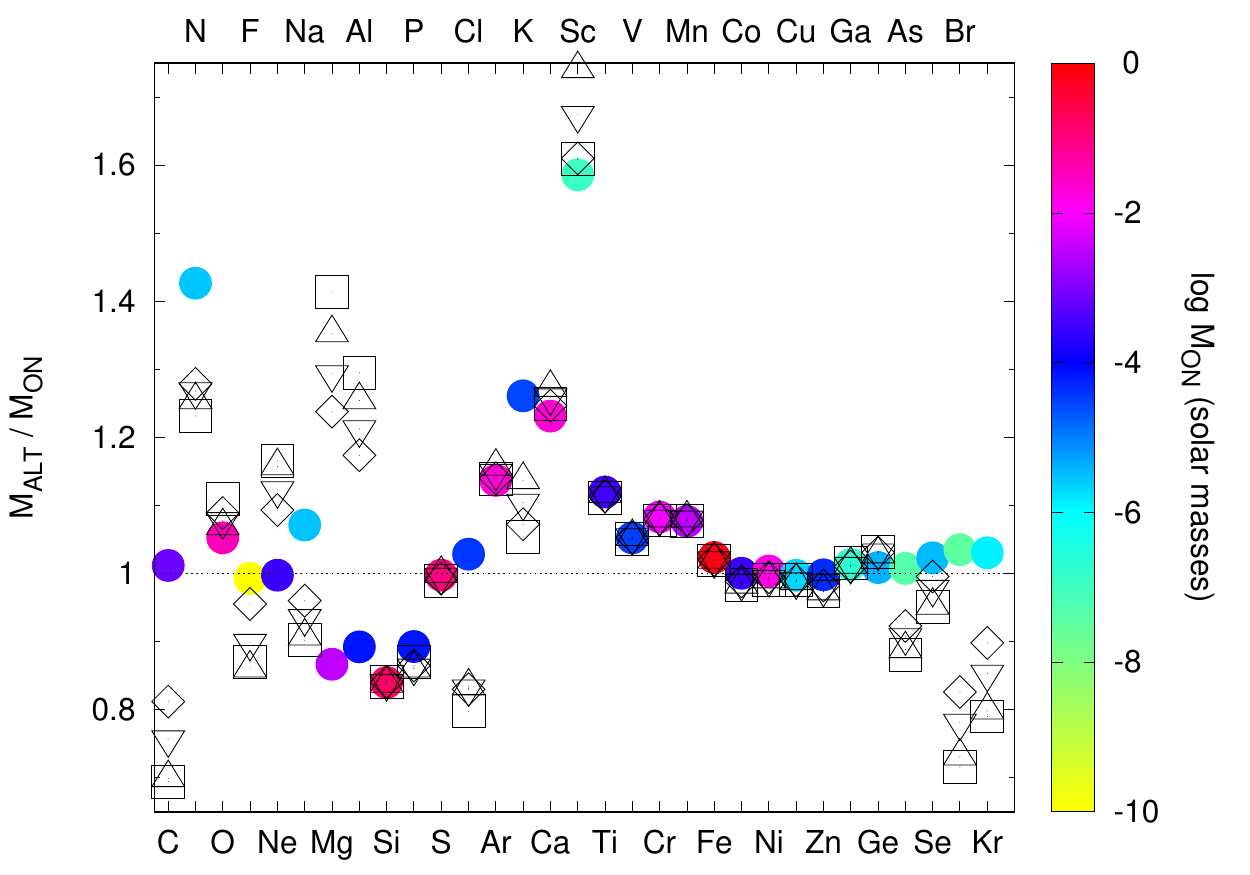}
    \caption{Relative change in the elemental yields derived from using alternative reaction rates 
(M$_\mathrm{ALT}$) instead of the standard ones (M$_\mathrm{ON}$, model 1p06\_Z9e-3\_std). The solid coloured 
circles belong to model 1p06\_Z9e-3\_$\xi_\mathrm{CO}$0p9, and the colour is assigned as a function of the 
yield of each element in model 1p06\_Z9e-3\_std. The rest of symbols belong to different combinations of 
modified rates of the reactions \isotope{12}{C}+\isotope{16}{O}, 
\isotope{12}{C}+\isotope{12}{C}, \isotope{16}{O}+\isotope{16}{O}, as well as 
\isotope{12}{C}$(\alpha,\gamma)$\isotope{16}{O} and its inverse, crafted to obtain similar ratios of calcium 
to sulfur yields (squares: model 1p06\_Z9e-3\_A, upward triangles: model 1p06\_Z9e-3\_B, downward triangles: model 1p06\_Z9e-3\_C, diamonds: model 1p06\_Z9e-3\_D,
see Section~\ref{s:losotros} for further details).
    \label{fig5}}
\end{figure}

Fig.~\ref{fig5} shows the relative changes introduced in the yields of the model with 
$M_\mathrm{WD}=1.06$~\msunb and $Z=0.009$ (coloured points), when $\xi_\mathrm{CO}=0.9$. The variations remain under 10\% for IGE and 
sulfur, while calcium, argon, and silicon are most affected among the elements with largest yields. The 
results for other explosion models (not shown in Fig.~\ref{fig5}) are similar. 

\begin{figure*}
   \includegraphics[width=\textwidth]{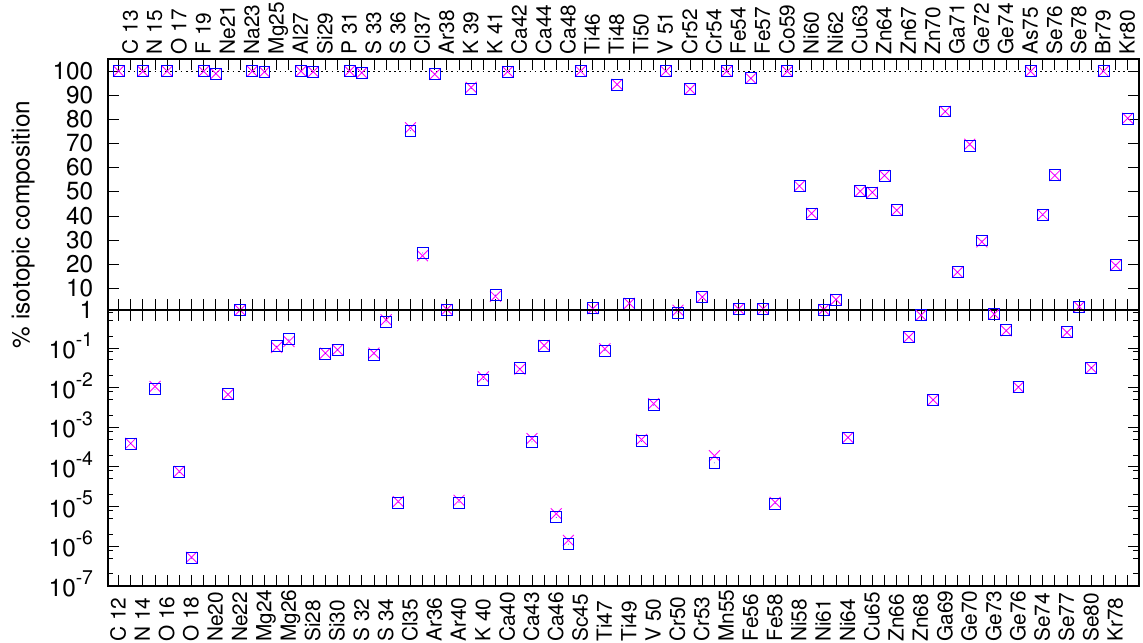}
    \caption{Percent isotopic contribution to the abundance of each element in models 1p06\_Z9e-3\_std 
(red crosses) and 1p06\_Z9e-3\_$\xi_\mathrm{CO}$0p9 (blue empty squares). This plot shows that the distribution of the yield of each element among its 
different stable isotopes is practically independent of the rate of the reaction 
\isotope{12}{C}+\isotope{16}{O}.
    \label{fig9}}
\end{figure*}

The isotopic contribution to the abundance of each element is not affected by the rate of the 
\isotope{12}{C}+\isotope{16}{O} reaction as much as the yield of the elements is, for the same model. Figure~\ref{fig9} shows the percent 
contribution of each isotope to the abundance of its element when the standard reaction rate is used and when 
$\xi_\mathrm{CO}=0.9$. In contrast with elemental mass yields, which change up to $40$\%, the percent 
contribution of each isotope is largely insensitive to the rate of \isotope{12}{C}+\isotope{16}{O}, with the 
exception of \isotope{54}{Cr}, whose contribution to chromium varies by as much as $\sim30$\%.

\subsection{Other reactions involving carbon and oxygen}\label{s:losotros}

The role of the reaction \isotope{12}{C}+\isotope{16}{O} in the synthesis of IME during {\sl explosive oxygen 
burning} was highlighted by \cite{1971woo} nearly five decades ago. They showed that the key point is 
the regulation of $\alpha$ particles produced per \isotope{28}{Si} nucleus, a process that is contributed by 
four reactions involving \isotope{12}{C} and \isotope{16}{O}: three fusion reactions, 
\isotope{12}{C}+\isotope{12}{C}, \isotope{12}{C}+\isotope{16}{O}, and \isotope{16}{O}+\isotope{16}{O}, plus 
the photodisintegration reaction \isotope{16}{O}+$\gamma\rightarrow$\isotope{12}{C}+$\alpha$. The last two 
reactions provide the main paths of destruction of oxygen. If 
\isotope{16}{O}+$\gamma\rightarrow$\isotope{12}{C}+$\alpha$ is followed by 
\isotope{12}{C}+\isotope{12}{C}$\rightarrow$\isotope{20}{Ne}+$\alpha$ and by 
\isotope{20}{Ne}+$\gamma\rightarrow$\isotope{16}{O}+$\alpha$, the net effect is the release of four $\alpha$ 
particles at the expense of one \isotope{16}{O} nucleus. Otherwise, the destruction of \isotope{16}{O}, 
either directly through \isotope{16}{O}+\isotope{16}{O} or indirectly by \isotope{12}{C}+\isotope{16}{O} just 
releases one $\alpha$ particle per each pair of \isotope{16}{O} nuclei consumed. Thus, two reactions 
contribute positively to a large $\alpha$ abundance, \isotope{16}{O}$\left(\gamma,\alpha\right)$
and \isotope{12}{C}+\isotope{12}{C}, and two contribute negatively, \isotope{12}{C}+\isotope{16}{O} and 
\isotope{16}{O}+\isotope{16}{O}. 
It is possible that a combination of relatively minor changes in the rates of these reactions might have the same net 
effect on the final Ca/S and Ar/S ratios as the scaling down of the \isotope{12}{C}+\isotope{16}{O} 
reaction rate by a factor ten. 

We have explored the results of modifying the four reaction rates by computing a set of 135 
additional versions of model 1p06\_Z9e-3. In each model, the three fusion reaction rates were 
multiplied by a random factor in the range $(0.5$--$1.5)$. To test the sensitivity to the rate of 
the photodisintegration reaction \isotope{16}{O}$\left(\gamma,\alpha\right)$\isotope{12}{C} (and its 
inverse), we used different rates \citep{1985cau,1988cau,1996buc,2012kat,2013xu} available in the JINA REACLIB 
database\footnote{https://groups.nscl.msu.edu/jina/reaclib/db/}.
The behaviour of the last reaction rate at temperatures in excess of 
$4\times10^9$~K is complicated and uncertain because of the possible contribution of high-energy resonances 
and cascade transitions \citep{2017deb}, and the use of different rates proposed in the literature is a 
convenient way to account for this uncertainty. 

\begin{figure}
   \includegraphics[width=\columnwidth]{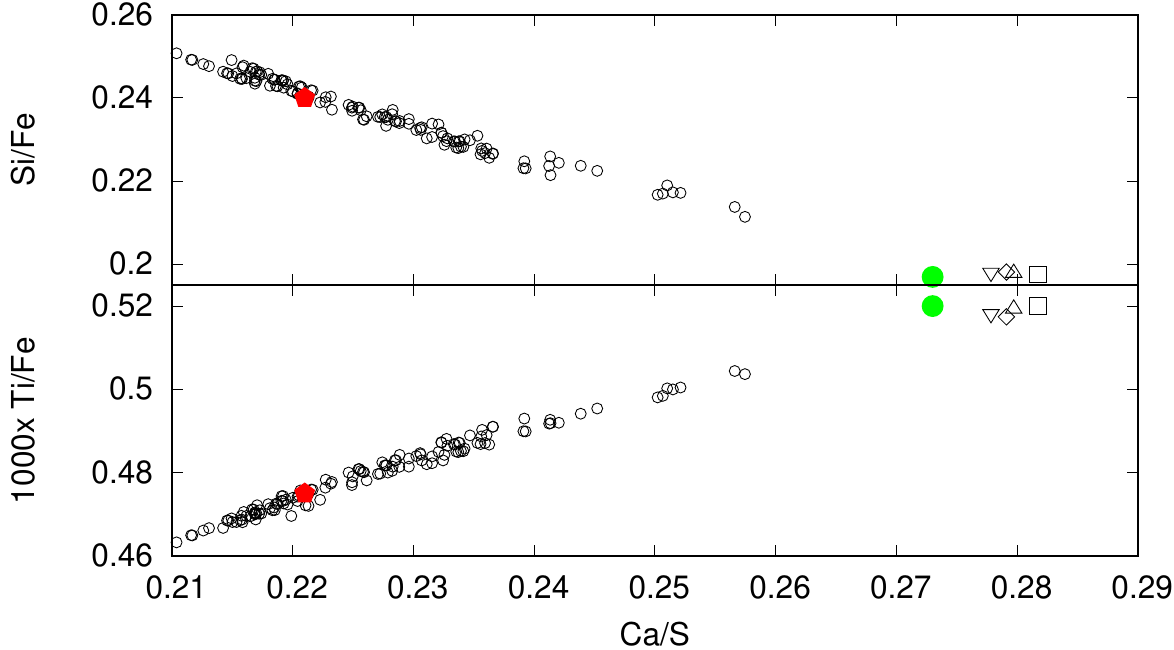}
    \caption{Variation in the ratios of silicon to iron and of titanium to iron vs the ratio of calcium to 
sulfur for a set of 139 models using random modifications of the reaction rates for 
\isotope{12}{C}+\isotope{16}{O}, 
\isotope{12}{C}+\isotope{12}{C}, \isotope{16}{O}+\isotope{16}{O}, and for
\isotope{12}{C}$(\alpha,\gamma)$\isotope{16}{O} and its inverse.
All models shown here are variants of the subCh model with $M_\mathrm{WD} = 1.06$~\msunb and $Z=0.009$. Red solid 
pentagons belong to model 1p06\_Z9e-3\_std, i.e. with all standard rates, while green solid circles belong to model 
1p06\_Z9e-3\_$\xi_\mathrm{CO}$0p9, which reproduces the Ca/S and Ar/S mass ratios in SNRs. Open circles belong to 
models obtained with the rates of the three fusion reactions above 
multiplied by a random factor in the range $(0.5$--$1.5)$, and the rate of photodisintegration of 
\isotope{16}{O} given by alternate recipes in the JINA REACLIB database. The rest of symbols belong to models 1p06\_Z9e-3\_A (squares),  1p06\_Z9e-3\_B (upward triangles), 1p06\_Z9e-3\_C (downward triangles), and 1p06\_Z9e-3\_D (diamonds), see text and Table~\ref{tabextra} for 
details.
    \label{fig10}}
\end{figure}

The resulting ratios of Ca/S, Si/Fe and Ti/Fe in these explosion models are shown by open circles in 
Fig.~\ref{fig10}, and cover a range of Ca/S between 0.21 and 0.26. These figures can be compared with the 
ratios obtained for model 1p06\_Z9e-3\_std, with all standard rates (red solid pentagons, located at Ca/S=0.22), and for 
model 1p06\_Z9e-3\_$\xi_\mathrm{CO}$0p9, with all reactions but \isotope{12}{C}+\isotope{16}{O} given by 
the standard rates (green solid circles, located at Ca/S=0.273).
The abundance ratios show an 
almost linear monotonic dependence on each other, irrespective of what reaction rate was modified and in 
which measure. In other words, the ratios of Si/Fe and Ti/Fe can be specified, for an explosion model like 
1p06\_Z9e-3, as a function of Ca/S, which is a quantity measurable in supernova remnants.

Motivated by these results, we have modified the four reaction rates involving 
carbon and oxygen in order to obtain Ca/S ratios similar to those for 
$\xi_\mathrm{CO}=0.9$, the value favoured by SNR measurements. For model 1p06\_Z9e-3\_$\xi_\mathrm{CO}$0p9, this
means Ca/S$\sim0.27-0.28$. For this experiment, we have had to change the four carbon and oxygen reaction rates
by more than the 50\% variations shown thus far.

Let us define a parameter to describe the scaling of each one of these reactions, similar to the definition of $\xi_\mathrm{CO}$ in Section~\ref{s:frame}. Thus, the rate of \isotope{12}{C}+\isotope{12}{C} is that from CF88 scaled by a factor 
$\left(1-\xi_\mathrm{CC}\right)$, the rate of \isotope{16}{O}+\isotope{16}{O} is that from CF88 scaled by a factor 
$\left(1-\xi_\mathrm{OO}\right)$ and, finally, the meaning of $\xi_\mathrm{CO}$ is the same as before. The parameters of the four models with these rates modified are given in Table~\ref{tabextra}. In all four models, the rates of \isotope{16}{O}$\left(\gamma,\alpha\right)$\isotope{12}{C} and its inverse were based on \cite{2012kat}.

\begin{table}
 \caption{Parameters of the four models with modified rates of reactions involving carbon and oxygen.}
 \label{tabextra}
 \begin{tabular}{@{}lrrr@{}}
 \hline\hline
 model name & $\xi_\mathrm{CC}$ & $\xi_\mathrm{OO}$ & $\xi_\mathrm{CO}$ \\
 \hline
 1p06\_Z9e-3\_A & -2.2 & 0.5 & 0.5 \\
 1p06\_Z9e-3\_B & -1.5 & 0.6 & 0.5 \\
 1p06\_Z9e-3\_C & -1.0 & 0.7 & 0.5 \\
 1p06\_Z9e-3\_D & -2.3 & 0.7 & 0.4 \\
 \hline\hline
 \end{tabular}
\end{table}

The point here is to explore to which extent fixing the Ca/S ratio in a given model, with respect to the variations in the rates 
of the carbon and oxygen reactions, is enough to determine the whole nucleosynthesis output 
or whether, on the contrary, the yields are sensitive to the precise reaction rates modified.
The results of the four modified models
are depicted in Fig.~\ref{fig10} as triangles, diamonds, and squares, close to the desired values, i.e. those of model 
1p06\_Z9e-3\_$\xi_\mathrm{CO}$0p9 (green solid circles). 

The full yields of the four crafted models can be seen 
as well in Fig.~\ref{fig5} as open symbols, to be compared with the ratios obtained for model 
1p06\_Z9e-3\_$\xi_\mathrm{CO}$0p9 (coloured solid circles). Besides IGE, whose yields we already have shown that are not affected by the
modified rates, the abundance ratios of elements 
which have an important contribution from explosive oxygen burning (see Figs.~\ref{fig7} and \ref{fig8}) are  
insensitive to the precise changes applied to the four reaction rates. 
To summarize, the combined variation of the rates in four key reactions involving \isotope{12}{C} and \isotope{16}{O} 
(\isotope{12}{C}+\isotope{16}{O}, 
\isotope{12}{C}+\isotope{12}{C}, \isotope{16}{O}+\isotope{16}{O}, and 
\isotope{16}{O}$+\gamma\rightarrow$\isotope{12}{C}$+\alpha$)
within their uncertainties can have the same effect than the suppression of the rate of the single reaction \isotope{12}{C}+\isotope{16}{O} 
by a factor $\xi_\mathrm{CO}=0.9$. In practice, this is true for the most important 
nucleosynthetic products of SN Ia, i.e. IGE and IME, with the exception of potassium and chlorine. These two elements are not a main product of explosive oxygen burning in our models, so their sensitivity to the four reaction rates is different from that of the products of this nucleosynthetic process (see Section~\ref{s:impact}).


One can wonder whether the exceptions to the above rule can provide a way to discriminate which ones of the four reactions should change with respect to their standard rates, and in which amount. This question is beyond the scope of the present paper and will need further investigation, but we can advance a few ideas. One possibility is to measure the ratio of two elements in SNRs, one which is sensitive to the changes in the four reaction rates and the other which is insensitive. For the first one, the first option that comes to mind is potassium, whose yield varies by $\sim30$\% in Fig.~\ref{fig5} and is produced in non-negligible amounts in the explosion. For the second one, a good choice could be calcium, whose yield range of variation is on the order of a few percent. 
Even though the changes in the yield of potassium are modest, future mid-term projected X-ray 
facilities with spectroscopic capabilities like {\sl XRISM} \citep[with $\sim5$~eV energy resolution and $\sim300$~cm$^2$ effective area;][]{2018tas,2018hit} 
and {\sl Athena} \citep[with $\sim2.5$~eV energy resolution and an effective area close to 1~m$^2$;][]{2013nan} 
should be able to discriminate them.

\section{Yields of SNR-calibrated SN Ia models}\label{s:yields}

In this Section, we give the final (after radioactive decays) elemental and isotopic yields of the 
SN Ia models obtained with $\xi_\mathrm{CO}=0.9$. We have shown that these models are representative of a class of models 
in which the rates of the four reactions involving \isotope{12}{C} and \isotope{16}{O} may change by 
different amounts but have in common the abundance ratio of calcium to sulfur and that of argon to sulfur, 
and are, in this sense, SNR-calibrated SN Ia models. 
In Section~\ref{s:radio}, we give the yields, in solar masses, of the most abundant radioactive isotopes with half-life longer than one day.
For completeness, we also give in Appendix~\ref{appa} the 
yields of the models using the standard set of reaction rates, \isotope{12}{C}+\isotope{16}{O} included.

\subsection{Chandrasekhar-mass models}\label{s:massive}

The yields belonging to the Chandra models are given in Table~\ref{tab:nuc-chan}. 
Rows starting by 
'elem' give the elemental yields of each model, in solar masses, where the element is identified by the 
atomic number and symbol.
Rows starting by 'isot' give the yield of each isotope in each model. In this case, the isotope is identified by the atomic number and symbol 
and by the baryon number.

\begin{table*}
\begin{minipage}{210mm}
\caption{Nucleosynthesis in Chandrasekhar-mass DDT models with $\xi_\mathrm{CO}=0.9$.}
\label{tab:nuc-chan}
\begin{tabular}{llllllllllll}
\hline
 \multicolumn{2}{l}{$\rho_\mathrm{DDT}$} &         1.2E+07 &  1.2E+07 &  1.2E+07 &  1.2E+07 &  1.2E+07 &  
1.6E+07 &  1.6E+07 &  1.6E+07 &  1.6E+07 &  1.6E+07 \\
 \multicolumn{2}{l}{$Z$} &              2.25E-4 &  2.25E-3 &  9.00E-3 &  2.25E-2 &  6.75E-2 &  2.25E-4 &  
2.25E-3 &  9.00E-3 &  2.25E-2 &  6.75E-2 \\
 elem &  2He  & 3.01E-04 & 2.87E-04 & 2.57E-04 & 1.61E-04 & 9.74E-06 & 3.08E-04 & 2.93E-04 & 2.62E-04 & 1.63E-04 & 9.43E-06 \\
 isot &  2He3 & 1.58E-12 & 1.56E-12 & 1.54E-12 & 1.42E-12 & 1.16E-12 & 1.58E-12 & 1.56E-12 & 1.54E-12 & 1.42E-12 & 1.16E-12 \\
 isot &  2He4 & 3.01E-04 & 2.87E-04 & 2.57E-04 & 1.61E-04 & 9.74E-06 & 3.07E-04 & 2.93E-04 & 2.62E-04 & 1.63E-04 & 9.43E-06 \\
 elem &  6C   & 4.94E-03 & 4.98E-03 & 4.90E-03 & 5.06E-03 & 5.21E-03 & 2.50E-03 & 2.47E-03 & 2.54E-03 & 2.64E-03 & 2.51E-03 \\
 isot &  6C12 & 4.94E-03 & 4.98E-03 & 4.90E-03 & 5.05E-03 & 5.21E-03 & 2.50E-03 & 2.47E-03 & 2.54E-03 & 2.64E-03 & 2.51E-03 \\
 isot &  6C13 & 1.94E-10 & 3.57E-09 & 1.26E-08 & 2.95E-08 & 7.43E-08 & 1.04E-10 & 1.64E-09 & 6.37E-09 & 1.57E-08 & 3.94E-08 \\
\hline
\end{tabular}
\vspace{-0.5cm}
\footnotetext{Sample of Table~\ref{tab:nuc-chan}, the full version is available online. The meaning of the 
columns is explained in the text.}
\end{minipage}
\end{table*}

\begin{table*}
\begin{minipage}{210mm}
\caption{Nucleosynthesis in sub-Chandrasekhar models with $\xi_\mathrm{CO}=0.9$.}
\label{tab:nuc-subch}
\begin{tabular}{llllllllllll}
\hline
 \multicolumn{2}{l}{$M_\mathrm{WD}$} &              0.88 &     0.88 &     0.88 &     0.88 &     0.88 &     
0.97 &     0.97 &     0.97 &     0.97 &     0.97 \\
 \multicolumn{2}{l}{$Z$} &              2.25E-4 &  2.25E-3 &  9.00E-3 &  2.25E-2 &  6.75E-2 &  
2.25E-4 &  2.25E-3 &  9.00E-3 &  2.25E-2 &  6.75E-2 \\
 elem &  2He  & 1.95E-07 & 1.94E-07 & 1.73E-07 & 1.30E-07 & 1.02E-08 & 2.21E-04 & 2.34E-04 & 2.67E-04 & 3.31E-04 & 4.90E-04 \\
 isot &  2He3 & 1.76E-16 & 1.76E-16 & 1.78E-16 & 1.83E-16 & 2.15E-16 & 6.37E-16 & 6.31E-16 & 6.14E-16 & 5.81E-16 & 4.83E-16 \\
 isot &  2He4 & 1.95E-07 & 1.94E-07 & 1.73E-07 & 1.29E-07 & 1.02E-08 & 2.20E-04 & 2.34E-04 & 2.67E-04 & 3.31E-04 & 4.90E-04 \\
 elem &  6C   & 4.04E-03 & 4.03E-03 & 3.96E-03 & 3.82E-03 & 3.35E-03 & 1.67E-03 & 1.66E-03 & 1.63E-03 & 1.57E-03 & 1.37E-03 \\
 isot &  6C12 & 4.04E-03 & 4.03E-03 & 3.96E-03 & 3.82E-03 & 3.34E-03 & 1.67E-03 & 1.66E-03 & 1.63E-03 & 1.56E-03 & 1.37E-03 \\
 isot &  6C13 & 2.20E-10 & 3.65E-09 & 1.21E-08 & 2.59E-08 & 5.61E-08 & 9.51E-11 & 1.52E-09 & 5.65E-09 & 1.28E-08 & 2.83E-08 \\
\hline
\end{tabular}
\vspace{-0.5cm}
\footnotetext{Sample of Table~\ref{tab:nuc-subch}, the full version is available online. The meaning of 
the columns is the same as in Table~\ref{tab:nuc-chan}.}
\end{minipage}
\end{table*}


\begin{figure}
   \includegraphics[width=\columnwidth]{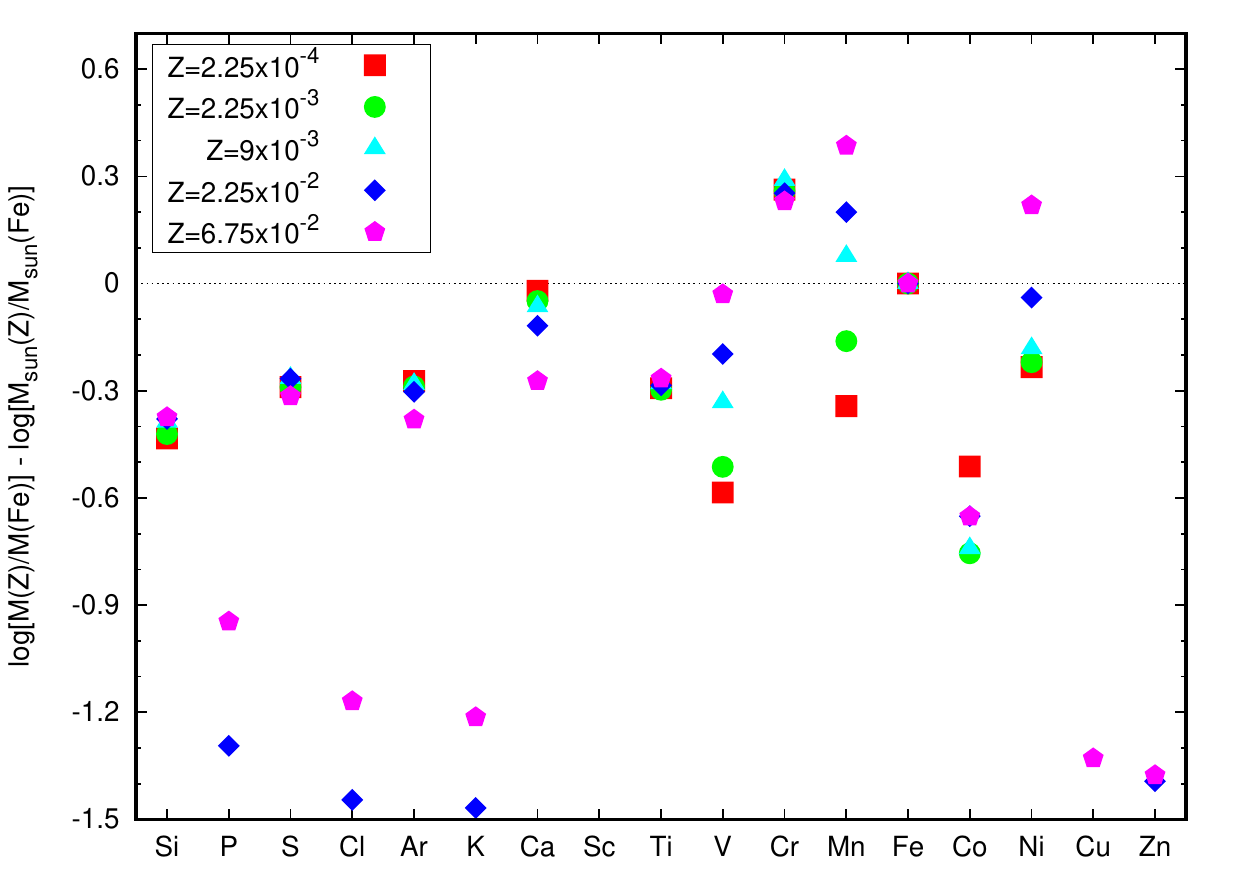}
    \caption{Variation of the elemental yields with respect to metallicity, for the Chandrasekhar-mass 
models with $\rho_\mathrm{DDT}=2.4\times10^7$~\gccb and $\xi_\mathrm{CO}=0.9$. The yields are normalized to Fe and to 
the solar abundances, in log scale.
    \label{fig11a}}
\end{figure}

\begin{figure}
   \includegraphics[width=\columnwidth]{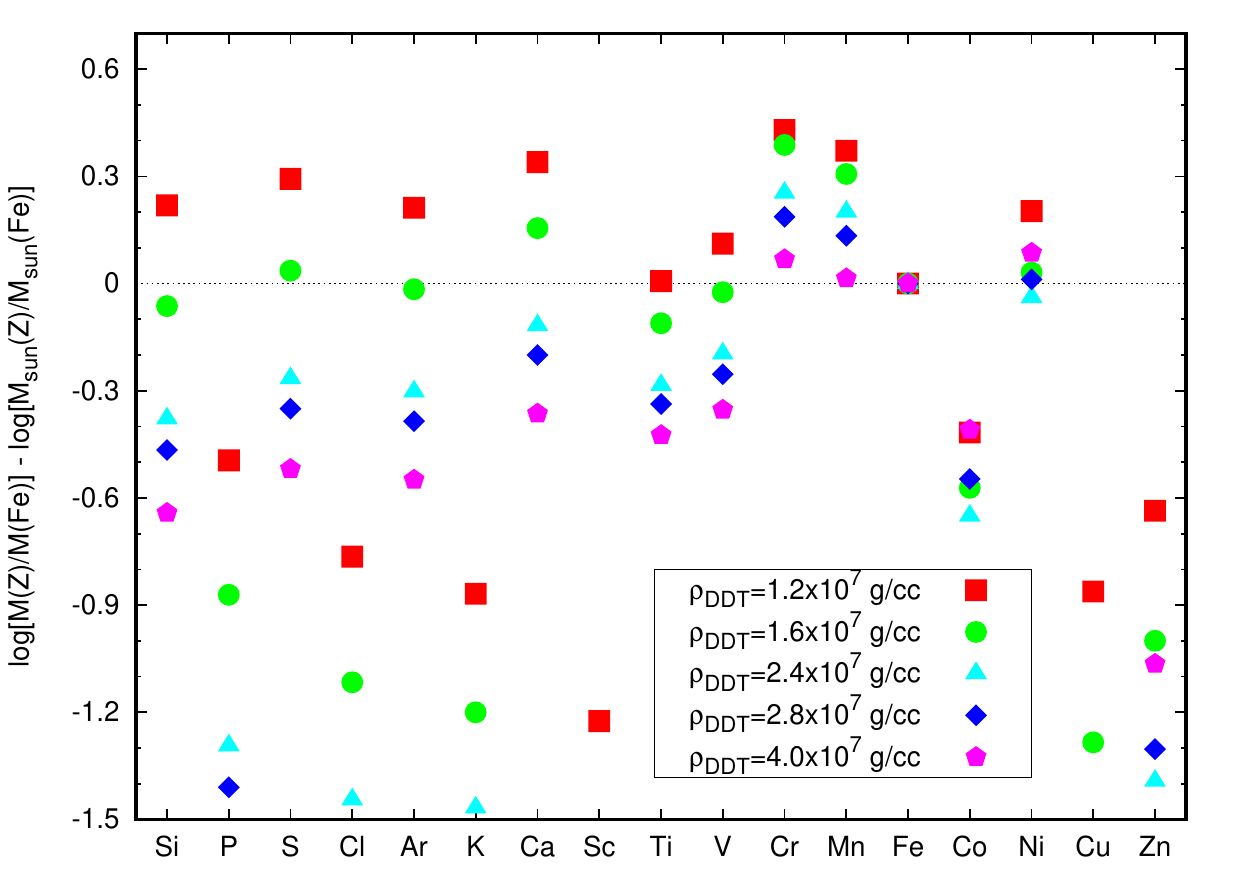}
    \caption{Variation of the elemental yields with respect to $\rho_\mathrm{DDT}$, for the 
models with $Z=0.0225$ and $\xi_\mathrm{CO}=0.9$. The yields are normalized to Fe and to 
the solar abundances, in log scale.
    \label{fig11b}}
\end{figure}


\begin{figure}
   \includegraphics[width=\columnwidth]{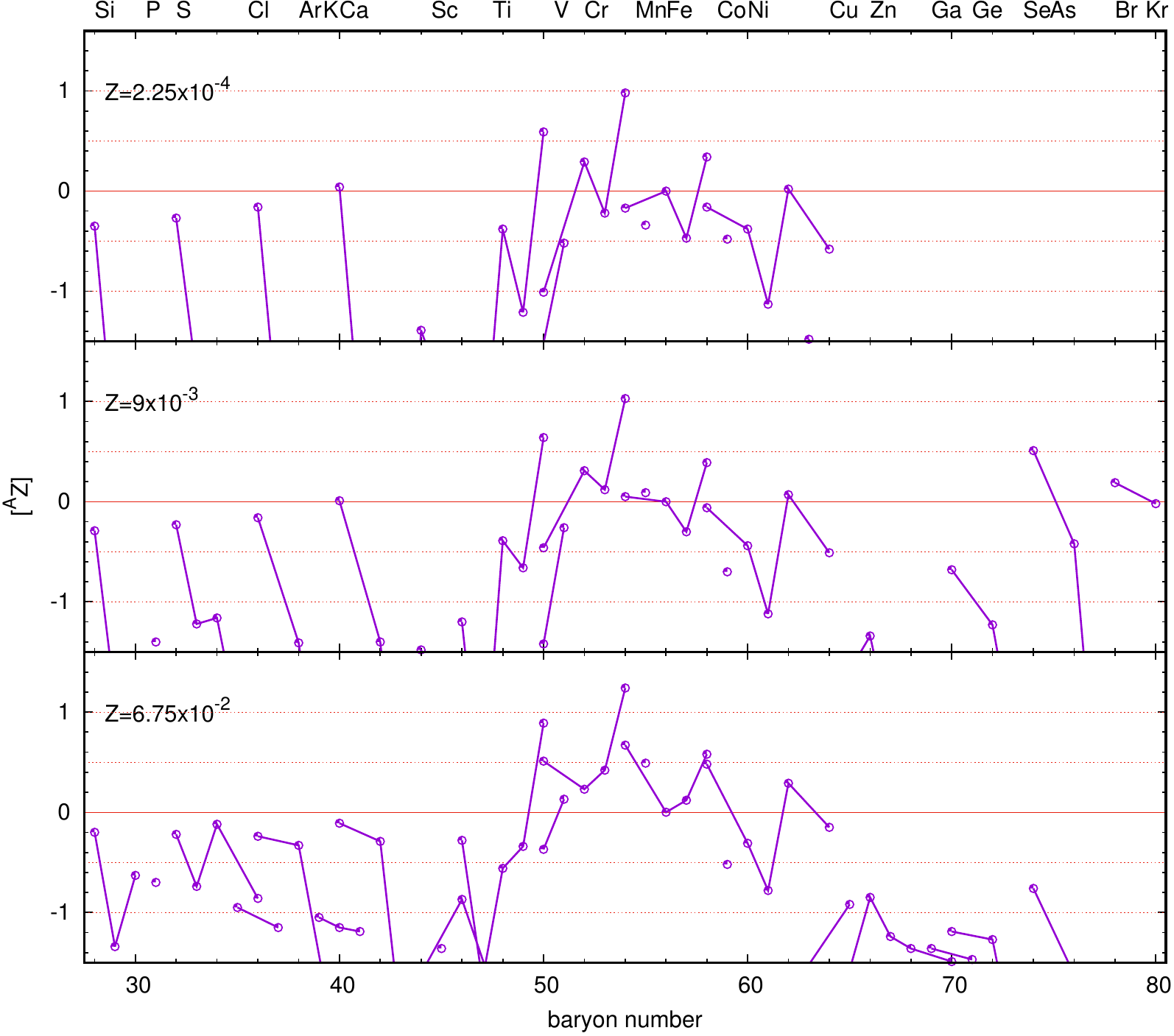}
    \caption{Variation of the isotopic yields with respect to metallicity, for the Chandrasekhar-mass 
models with $\rho_\mathrm{DDT}=2.4\times10^7$~\gccb and $\xi_\mathrm{CO}=0.9$. The yields are normalized to 
\isotope{56}{Fe} and to the solar abundances, in log scale. The approximate locus of each element is labelled 
on the top axis.
    \label{fig12a}}
\end{figure}

\begin{figure}
   \includegraphics[width=\columnwidth]{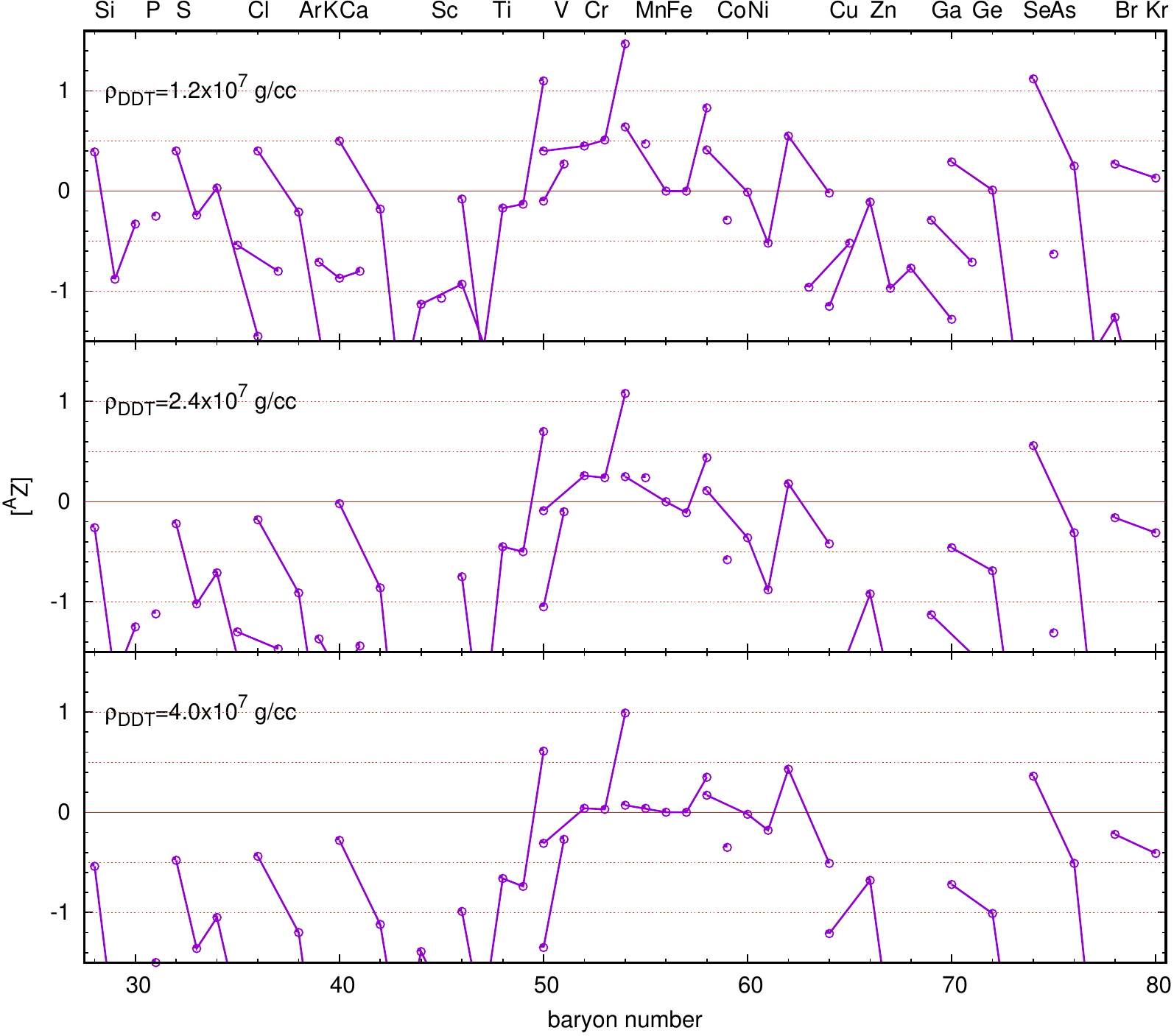}
    \caption{Variation of the isotopic yields with respect to $\rho_\mathrm{DDT}$, for the 
models with $Z=0.0225$ and $\xi_\mathrm{CO}=0.9$. The yields are normalized to 
\isotope{56}{Fe} and to the solar abundances, in log scale. The approximate locus of each element is labelled 
on the top axis.
    \label{fig12b}}
\end{figure}

The behaviour of the yields of the most abundant explosion products with respect to 
variations of the parameters of the Chandra models, $\rho_\mathrm{DDT}$ and $Z$, is illustrated in 
Figs.~\ref{fig11a} and \ref{fig11b} (elements) and \ref{fig12a} and \ref{fig12b} (isotopes). 
It should be noted that the abundance scales are normalized to 
iron (and to the solar ratio), which means that a change in the relative abundance with, for instance, 
 $\rho_\mathrm{DDT}$ may be due to a varying iron yield, to a change in the element yield, or to both. In these 
figures, 
the dependence on metallicity is illustrated taking as a reference the models with $\rho_\mathrm{DDT} = 
2.4\times10^7$~\gcc, 
whose $^{56}$Ni yields are representative of normal-luminosity SN Ia. 

The most remarkable feature in Fig.~\ref{fig11b} is that the ratios of all elements, with respect to 
iron, decrease monotonously 
with increasing deflagration-to-detonation transition density. It is a consequence of the increasing mass 
of IGE with 
increasing  $\rho_\mathrm{DDT}$, at given metallicity. Furthermore, most of the additional mass that is 
processed by nuclear reactions above the DDT layer 
is made of iron. Thus, all elements but iron show similar dependence on  $\rho_\mathrm{DDT}$. 
Manganese, chromium, and nickel are produced in almost solar proportions with respect to iron in the most 
luminous models, i.e. those with the largest \isotope{56}{Ni} 
yields. The same is true for titanium and vanadium in the most subluminous models. The yields of a few elements, such as zinc, 
do not change monotonously with  $\rho_\mathrm{DDT}$. 
The only intermediate-mass elements produced in abundance are the even-atomic number ones: silicon, sulfur, 
argon, and calcium. 
They are even overproduced, with respect to iron and the solar proportions, in the most subluminous models. On 
the other hand, scandium is
under-abundant in all present models.

The variation of the mass ratios of the different elements with respect to metallicity does not behave as homogeneously as with respect to  
$\rho_\mathrm{DDT}$. As can be appreciated in 
Fig.~\ref{fig11a}, by increasing $Z$ one obtains larger yields of the odd-atomic number elements, with the 
exception of cobalt, while most of the even-atomic number ones remain 
largely unaffected, with the exception of nickel. Thus, the mass ratio of odd-to-even atomic number elements 
is, in general, sensitive to the progenitor
metallicity, a result that has been already used to measure the metallicity of supernova progenitors through 
the ejecta abundances in supernova remnants 
\citep[e.g.][]{2008bad}. The mass ratio of potassium to calcium 
is affected by the progenitor metallicity, so one has to fix the last in order to use this mass ratio to get 
insight on the rate of the reactions involving 
carbon and oxygen, as explained in the previous section. 

Isotopic mass ratios can not be directly measured in supernova remnants, but they are necessary 
ingredients in galactic chemical evolution models. 
In Figs.~\ref{fig12a} and \ref{fig12b}, we show the isotopic ratios for baryon number $A\ge28$. Most isotopes from chromium to nickel are produced in almost solar proportions in the 
most luminous models (Fig.~\ref{fig12b}).
The exceptions are $^{50}$Ti, $^{54}$Cr, $^{58}$Fe, and, to a lesser extent, $^{62}$Ni, which are overproduced 
in all models. This is due to the relatively
high central density of the explosion models, $\rho_\mathrm{c}=3\times10^9$~\gcc, which implies that matter is 
efficiently neutronized
by electron captures on IGE isotopes in nuclear statistical equilibrium, shortly after central incineration, during 
the deflagration phase of the explosion. The
isotope $^{74}$Se is overproduced at all $\rho_\mathrm{DDT}$ in models with metallicity ranging from slightly 
subsolar to slightly supersolar (Fig.~\ref{fig12a}).
In the same metallicity range, the krypton isotopes $^{78}$Kr and $^{80}$Kr are produced in almost solar 
proportions with respect to $^{56}$Fe.

The isotopes that are most sensitive to $Z$ are 
$^{46}$Ti, $^{49}$Ti, $^{50}$V, $^{51}$V, $^{50}$Cr, $^{53}$Cr, $^{55}$Mn, $^{54}$Fe, $^{57}$Fe, and 
$^{58}$Ni. We note that odd-baryon number and
even-baryon number isotopes are equally represented in this list. Thus, the dependence of these isotope 
yields with respect to metallicity should be taken into account in
chemical evolution models. 

\subsection{Sub-Chandrasekhar models}\label{s:subch}


\begin{figure}
   \includegraphics[width=\columnwidth]{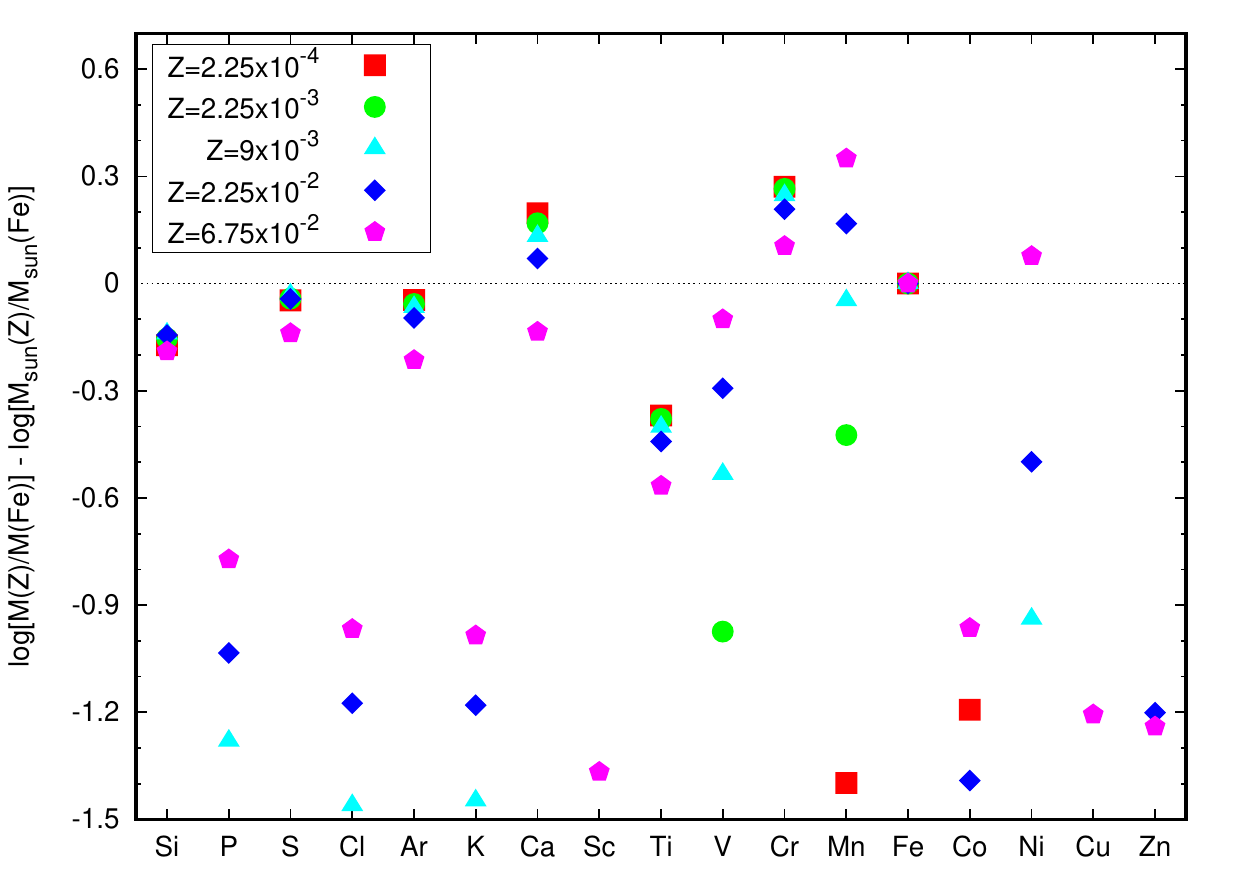}
    \caption{Variation of the elemental yields with respect to metallicity, for the sub-Chandrasekhar models 
with $M_\mathrm{WD}=0.97$~\msunb and $\xi_\mathrm{CO}=0.9$. 
    \label{fig13a}}
\end{figure}

\begin{figure}
   \includegraphics[width=\columnwidth]{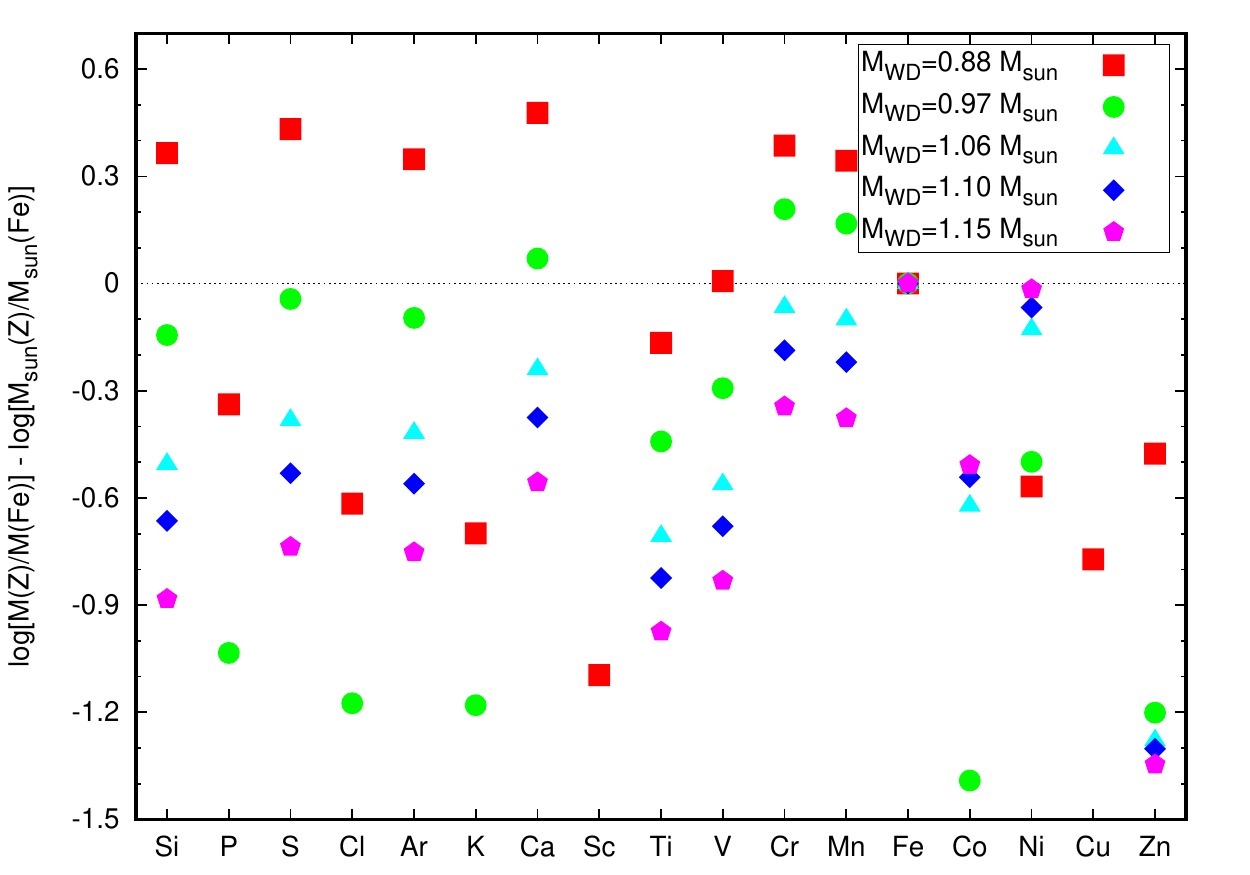}
    \caption{Variation of the elemental yields with respect to the WD mass, for the sub-Chandrasekhar models 
with $Z=0.0225$ and $\xi_\mathrm{CO}=0.9$. 
    \label{fig13b}}
\end{figure}


\begin{figure}
   \includegraphics[width=\columnwidth]{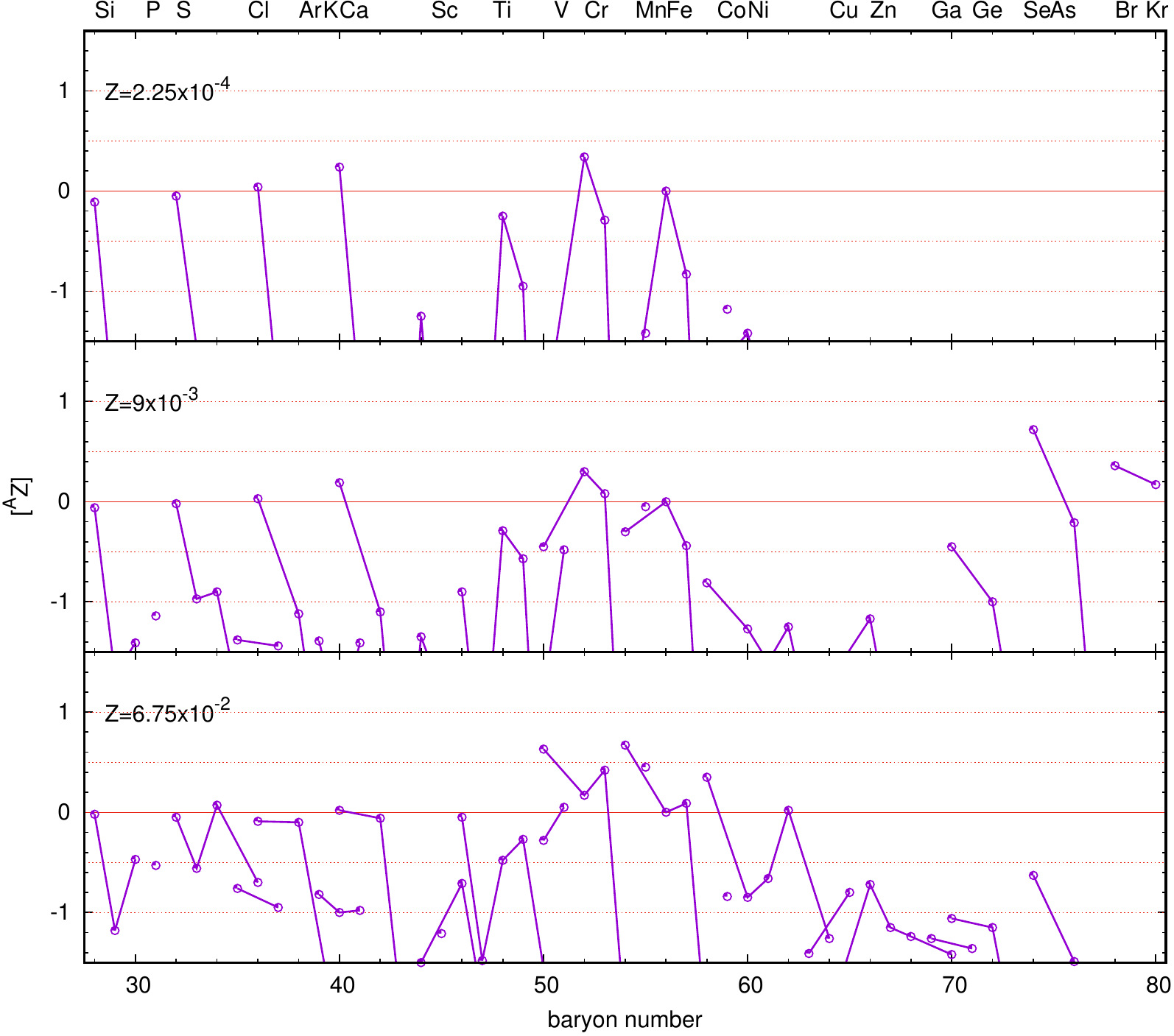}
    \caption{Variation of the isotopic yields with respect to metallicity, for the sub-Chandrasekhar models 
with $M_\mathrm{WD}=0.97$~\msunb and $\xi_\mathrm{CO}=0.9$.
    \label{fig14a}}
\end{figure}

\begin{figure}
   \includegraphics[width=\columnwidth]{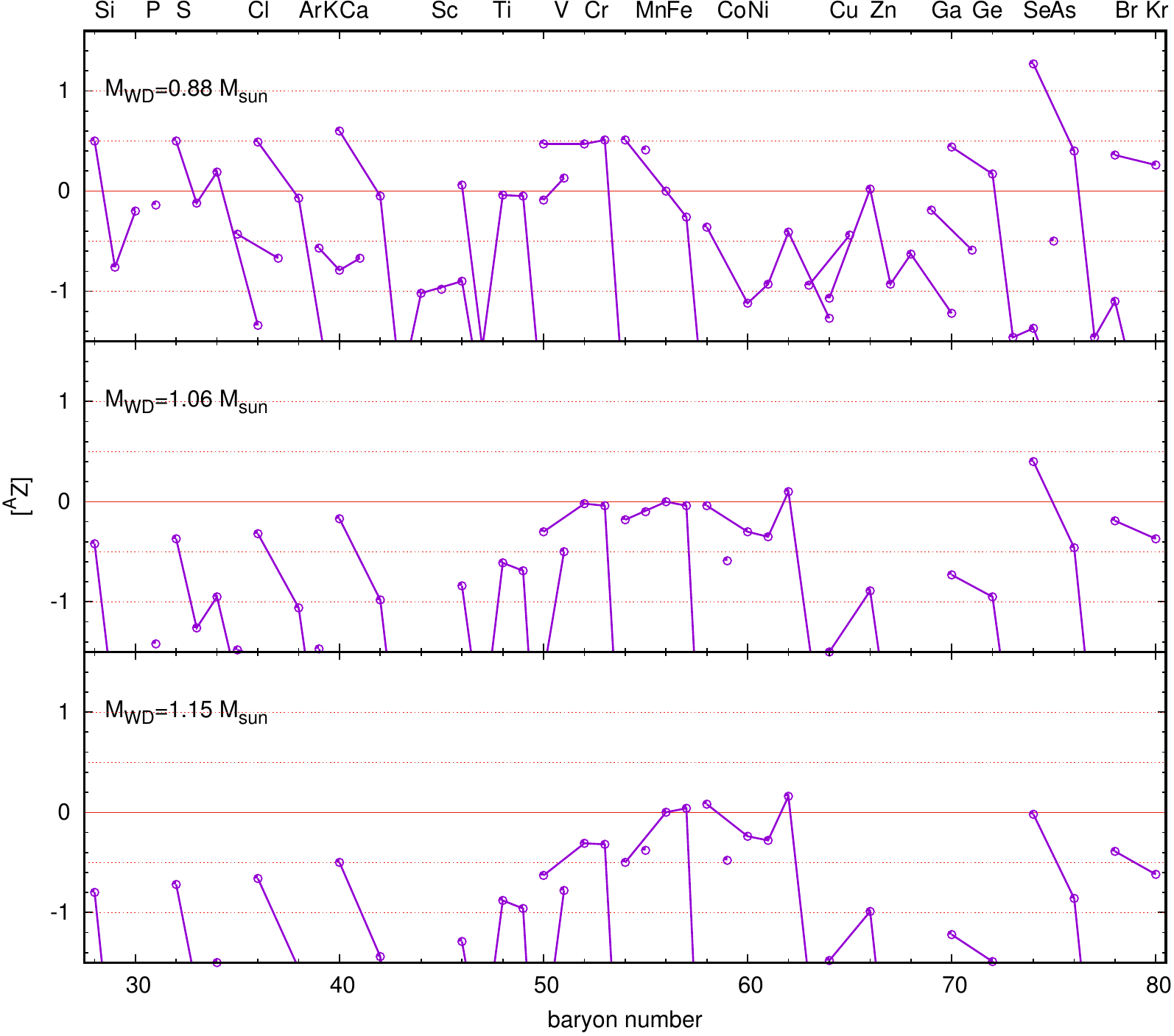}
    \caption{Variation of the isotopic yields with respect to the WD mass, for the sub-Chandrasekhar models 
with $Z=0.0225$ and $\xi_\mathrm{CO}=0.9$.
    \label{fig14b}}
\end{figure}

The yields belonging to the sub Chandrasekhar-mass models are given in Table~\ref{tab:nuc-subch}, and 
illustrated in 
Figs.~\ref{fig13a} to \ref{fig14b}.  In these figures, 
the dependence on metallicity is illustrated taking as a reference the models with $M_\mathrm{WD} = 
0.97$~\msun, 
whose $^{56}$Ni yields are representative of slightly subluminos SN Ia. 

The trends with progenitor metallicity of the yields of elements  and isotopes in sub-Chandrasekhar models 
(Figs.~\ref{fig13a} and \ref{fig14a}) resemble those in Chandrasekhar models, discussed in 
Section~\ref{s:massive}, the main
difference being that in subCh models the nickel isotopic yields are strongly metallicity dependent. 

The WD mass in subCh models plays a role similar to that of the deflagration-to-detonation transition density 
in Chandra models, 
in the sense that it is the leading parameter determining the ejected mass of $^{56}$Ni and, hence, the 
luminosity of the supernova. 
Here (Fig.~\ref{fig13b}), as in Chandra models, the ratios of most elements with respect to iron 
decrease monotonously with increasing 
iron yield. However, there are two notable exceptions, cobalt and nickel, whose yields increase further than 
that of iron with increasing WD mass.

Most isotopes from chromium to nickel are produced in almost solar proportions in the normal luminosity models 
(Fig.~\ref{fig14b}).
In contrast with Chandra models, $^{50}$Ti, $^{54}$Cr, and $^{58}$Fe are underproduced in all models, which 
may help to compensate for their
overproduction in Chandra models if both explosion scenarios contribute in similar proportions to SN Ia.

\subsection{Radioactivities}\label{s:radio}

Tables~\ref{tab:rad-chan} and \ref{tab:rad-subch} give the yields of the most abundant radioactive isotopes with half-life longer than one day. We have included all the isotopes whose yield in any one of the models presented in this work is larger than $10^{-6}$~\msun.

Table~\ref{tab:isorad} gives the maximum yield of these radioactive isotopes in all our models. The yields of typical targets of $\gamma$-ray observations of supernova remnants, e.g. \isotope{26}{Al} and \isotope{44}{Ti}, are pretty small, on the order of $10^{-5}$ -- $10^{-6}$~\msun, while other isotopes, e.g. \isotope{59}{Ni}, are produced in interesting amounts.

\begin{table*}
\begin{minipage}{210mm}
\caption{Radioactivities with half-life longer than one day in Chandrasekhar-mass DDT models with $\xi_\mathrm{CO}=0.9$.}
\label{tab:rad-chan}
\begin{tabular}{lllllllllll}
\hline
 \multicolumn{1}{l}{$\rho_\mathrm{DDT}$} &         1.2E+07 &  1.2E+07 &  1.2E+07 &  1.2E+07 &  1.2E+07 &  
1.6E+07 &  1.6E+07 &  1.6E+07 &  1.6E+07 &  1.6E+07 \\
 \multicolumn{1}{l}{$Z$} &              2.25E-4 &  2.25E-3 &  9.00E-3 &  2.25E-2 &  6.75E-2 &  2.25E-4 &  
2.25E-3 &  9.00E-3 &  2.25E-2 &  6.75E-2 \\
 Al26 &   1.07E-07 & 2.87E-07 & 3.99E-07 & 6.63E-07 & 1.18E-06 & 8.41E-08 & 1.48E-07 & 2.07E-07 & 3.31E-07 & 5.10E-07 \\
 P 32 &   1.97E-10 & 2.48E-08 & 1.36E-07 & 5.23E-07 & 4.89E-06 & 1.28E-10 & 1.66E-08 & 8.86E-08 & 3.28E-07 & 2.71E-06 \\
 P 33 &   8.11E-11 & 1.86E-08 & 1.26E-07 & 4.90E-07 & 3.81E-06 & 4.74E-11 & 1.11E-08 & 7.68E-08 & 3.00E-07 & 2.19E-06 \\
 S 35 &   2.46E-11 & 6.29E-09 & 7.54E-08 & 7.20E-07 & 1.11E-05 & 1.47E-11 & 4.07E-09 & 4.86E-08 & 4.48E-07 & 6.34E-06 \\
 Ar37 &   6.52E-06 & 1.17E-05 & 2.47E-05 & 4.01E-05 & 6.09E-05 & 4.55E-06 & 8.71E-06 & 1.78E-05 & 2.87E-05 & 4.48E-05 \\
 Ca41 &   9.76E-07 & 2.19E-06 & 5.51E-06 & 9.45E-06 & 1.31E-05 & 8.78E-07 & 1.77E-06 & 4.18E-06 & 6.92E-06 & 9.82E-06 \\
\hline
\end{tabular}
\vspace{-0.5cm}
\footnotetext{Sample of Table~\ref{tab:rad-chan}, the full version is available online. The meaning of the 
columns is explained in the text.}
\end{minipage}
\end{table*}

\begin{table*}
\begin{minipage}{210mm}
\caption{Radioactivities with half-life longer than one day in sub-Chandrasekhar models with $\xi_\mathrm{CO}=0.9$.}
\label{tab:rad-subch}
\begin{tabular}{lllllllllll}
\hline
 \multicolumn{1}{l}{$M_\mathrm{WD}$} &              0.88 &     0.88 &     0.88 &     0.88 &     0.88 &     
0.97 &     0.97 &     0.97 &     0.97 &     0.97 \\
 \multicolumn{1}{l}{$Z$} &              2.25E-4 &  2.25E-3 &  9.00E-3 &  2.25E-2 &  6.75E-2 &  2.25E-4 &  
2.25E-3 &  9.00E-3 &  2.25E-2 &  6.75E-2 \\
 Al26 &   7.78E-08 & 2.03E-07 & 2.80E-07 & 4.27E-07 & 6.16E-07 & 4.06E-08 & 7.53E-08 & 1.02E-07 & 1.46E-07 & 1.92E-07 \\
 P 32 &   1.47E-10 & 1.96E-08 & 1.06E-07 & 3.82E-07 & 2.98E-06 & 7.04E-11 & 9.88E-09 & 4.99E-08 & 1.68E-07 & 1.18E-06 \\
 P 33 &   5.44E-11 & 1.46E-08 & 9.79E-08 & 3.61E-07 & 2.42E-06 & 2.68E-11 & 6.24E-09 & 4.03E-08 & 1.48E-07 & 9.83E-07 \\
 S 35 &   1.83E-11 & 4.98E-09 & 5.64E-08 & 4.94E-07 & 6.51E-06 & 8.28E-12 & 2.70E-09 & 2.88E-08 & 2.30E-07 & 2.77E-06 \\
 Ar37 &   5.63E-06 & 9.53E-06 & 1.93E-05 & 2.97E-05 & 4.42E-05 & 4.24E-06 & 7.82E-06 & 1.46E-05 & 2.29E-05 & 3.47E-05 \\
 Ca41 &   8.23E-07 & 1.88E-06 & 4.49E-06 & 7.07E-06 & 9.58E-06 & 9.39E-07 & 1.79E-06 & 3.61E-06 & 5.71E-06 & 7.96E-06 \\
\hline
\end{tabular}
\vspace{-0.5cm}
\footnotetext{Sample of Table~\ref{tab:rad-subch}, the full version is available online. The meaning of the 
columns is explained in the text.}
\end{minipage}
\end{table*}

\begin{table}
 \caption{Maximum yield, in solar masses, of the radioactive isotopes in all our models.}
 \label{tab:isorad}
 \begin{tabular}{lr@{\hspace{2truecm}}lr}
 \hline
 Al26 & $1.7\times10^{-6}$ & Mn54 & $1.2\times10^{-5}$  \\
 P 32 & $4.9\times10^{-6}$ & Fe55 & $2.0\times10^{-2}$  \\
 P 33 & $3.8\times10^{-6}$ & Fe59 & $1.2\times10^{-5}$  \\
 S 35 & $1.1\times10^{-5}$ & Fe60 & $1.4\times10^{-4}$  \\
 Ar37 & $6.5\times10^{-5}$ & Co56 & $1.7\times10^{-4}$  \\
 Ca41 & $1.3\times10^{-5}$ & Co57 & $8.4\times10^{-4}$  \\
 Ti44 & $5.1\times10^{-5}$ & Co58 & $4.8\times10^{-6}$  \\
 V 48 & $6.8\times10^{-4}$ & Co60 & $7.4\times10^{-6}$  \\
 V 49 & $4.0\times10^{-5}$ & Ni56 & $9.3\times10^{-1}$  \\
 Cr51 & $1.8\times10^{-4}$ & Ni57 & $3.4\times10^{-2}$  \\
 Mn52 & $1.6\times10^{-2}$ & Ni59 & $1.6\times10^{-3}$  \\
 Mn53 & $2.2\times10^{-3}$ & Zn65 & $2.7\times10^{-6}$  \\
 \hline
 \end{tabular}
\end{table}

\section{Conclusions}\label{s:conclu}

We have computed the nucleosynthesis and hydrodynamics output for spherically symmetric models of SN Ia
belonging to two explosion paradigms: the delayed-detonation of a Chandrasekhar-mass WD, and the pure central 
detonation of a sub-Chandrasekhar mass WD. Our models differ from existing 
compilations of SN Ia nucleosynthesis in two aspects. First, we have computed the nucleosynthesis 
using a large nuclear network of up to 722 nuclides in
the hydrodynamics code, instead of post-processing the output. 
Second, we introduce the concept of SNR-calibrated SN Ia nucleosynthesis models.

Since our models are one-dimensional they cannot account for the effect of hydrodynamical instabilities and turbulence that are regularly found in multidimensional simulations of SN Ia \citep[e.g.][]{2004ple,2006rop,2006bra,2007kasb}. In spite of these shortcomings, the results of many one-dimensional models compare well with observations of both SN Ia and their remnants \citep{1996hoe,1997nug,2006bad,2013blo,2017hoe,2018mar}, many SN Ia show a high level of stratification, in better agreement with one-dimensional models than multidimensional ones \citep[e.g.][]{2010tanb}, and many remnants display geometries close to spherical symmetry \citep{2011lop}. However, the reader should be aware of the limitations of the one-dimensional hydrodynamical approach.

We have established that the combined variation of the rates in four key reactions involving \isotope{12}{C} and \isotope{16}{O} 
(\isotope{12}{C}+\isotope{16}{O}, 
\isotope{12}{C}+\isotope{12}{C}, \isotope{16}{O}+\isotope{16}{O}, and 
\isotope{16}{O}$+\gamma\rightarrow$\isotope{12}{C}$+\alpha$)
within the uncertainties has the same effect than the suppression of the rate of the single reaction \isotope{12}{C}+\isotope{16}{O} 
by a factor $\xi_\mathrm{CO}=0.9$. 
For the sake of simplicity, we adopt this single change as representative of the changes that are needed to reconcile the 
nucleosynthetic yields of SN Ia models with the Ca/S and Ar/S mass ratios measured in Galactic and Magellanic Cloud Type Ia SNRs 
\citep{2017mar}, and call this family of modified SN Ia models ``SNR-calibrated SN Ia models''.  

For all models, we have computed the hydrodynamics and nucleosynthesis starting from WDs with metallicities in 
the range from $Z=0.000225$ to $Z=0.0675$. For Chandrasekhar-mass models we have computed explosions 
with deflagration-to-detonation transition densities ranging from $1.2\times10^7$~\gccb to 
$4.0\times10^7$~\gcc, while for sub-Chandrasekhar mass models we have allowed the mass of the exploding WD to 
vary between $0.88$~\msunb and $1.15$~\msun. The explosions of Chandrasekhar-mass WDs with $\rho_\mathrm{DDT}=2.4\times10^7$~\gccb and of 
sub-Chandrasekhar mass WDs with $M_\mathrm{WD}=1.06$~\msunb should be representative of normal-luminosity 
SN Ia, since they produce between $0.55$~\msunb and $0.75$~\msunb of \isotope{56}{Ni}.

There is a remarkable difference between Chandrasekhar-mass and sub-Chandrasekhar mass models. In the first ones, the increment in the IGE 
yields with $\rho_\mathrm{DDT}$ comes predominantly in the form of iron. In the second ones, this increment of IGE with $M_\mathrm{WD}$ is contributed 
significantly by cobalt and nickel. We also note that the even-atomic number IMEs are overproduced with 
respect to iron in the models that produce the less \isotope{56}{Ni} (hence, the most sub-luminous ones), for 
all WD masses. 

The neutron-rich isotopes \isotope{50}{Ti}, \isotope{54}{Cr}, and \isotope{58}{Fe} are overproduced 
with respect to \isotope{56}{Fe} in the Chandrasekhar-mass models because of the high initial central 
density. On the other hand, the same isotopes are underproduced in all sub-Chandrasekhar mass 
models, so a combination of explosions of all masses might be able to produce the proportions of 
these isotopes in the Solar System.

We notice an important production of the isotopes \isotope{74,76}{Se} and \isotope{78,80}{Kr}, 
which are subdominant with respect to the selenium and krypton isotopic composition in the Solar System. 
All these isotopes are less neutron-rich than the most abundant ones in the Solar System, and are produced in the outermost 
shells of the exploding WD.

Future X-ray facilities like {\sl XRISM} and {\sl Athena} may be able to discriminate models with respect to the rates of  
the aforementioned four reactions involving carbon and oxygen.
In this respect, the mass ratio of potassium to calcium looks as the most promising observational target, 
although this point is in need of further investigation. 
One possible strategy would be to 
fix the properties of the ISM and, if unknown, the age of observed supernova remnants through the most prominent lines in the SNR X-ray spectra and,
as well, fix the best explosion model describing the spectra. 
Thereafter, the mass ratio K/Ca would provide the best estimate of the involved reaction rates, since the yield of K is sensitive to their values.
The results would have to be cross-checked for as many SNRs as possible, 
in order to provide a statistically convincing constraint on the relevant nuclear physics. 

\section*{Acknowledgements}

We thank the anonymous referee for a very detailed and constructive report, which has contributed to improve the presentation of this work.
E.B. acknowledges founding from the MINECO-FEDER grant AYA2015-63588-P; H.M.-R. acknowledges support from 
NASA ADAP grant NNX15AM03G S01, a PITT PACC, and a Zaccheus Daniel Predoctoral Fellowship. E.B. wishes to
dedicate this work to the loving memory of Wilma.

\bibliographystyle{mnras} 

\appendix

\section{Yields of models with the standard $^{12}$C+$^{16}$O reaction rate}\label{appa}

In this Section, we give the final (after radioactive decays) elemental and isotopic yields of the 
SN Ia models that use the standard set of reaction rates, $^{12}$C+$^{16}$O included (Tables~\ref{tab:nuc-chan-std} and \ref{tab:nuc-subch-std}). 
We give as well the yields of the most abundant radioactive isotopes with half-life longer than one day (Tables~\ref{tab:rad-chan-std} and \ref{tab:rad-subch-std}).

\begin{table*}
\begin{minipage}{210mm}
\caption{Nucleosynthesis in Chandrasekhar-mass DDT models with standard $^{12}$C+$^{16}$O reaction rate.}
\label{tab:nuc-chan-std}
\begin{tabular}{llllllllllll}
\hline
 \multicolumn{2}{l}{$\rho_\mathrm{DDT}$} &         1.2E+07 &  1.2E+07 &  1.2E+07 &  1.2E+07 &  1.2E+07 &  
1.6E+07 &  1.6E+07 &  1.6E+07 &  1.6E+07 &  1.6E+07 \\
 \multicolumn{2}{l}{$Z$} &              2.25E-4 &  2.25E-3 &  9.00E-3 &  2.25E-2 &  6.75E-2 &  2.25E-4 &  
2.25E-3 &  9.00E-3 &  2.25E-2 &  6.75E-2 \\
 elem &  2He  & 1.31E-04 & 1.22E-04 & 1.00E-04 & 4.64E-05 & 3.05E-08 & 1.32E-04 & 1.23E-04 & 1.01E-04 & 4.63E-05 & 3.18E-08 \\
 isot &  2He3 & 1.80E-12 & 1.78E-12 & 1.76E-12 & 1.62E-12 & 1.33E-12 & 1.80E-12 & 1.78E-12 & 1.76E-12 & 1.62E-12 & 1.33E-12 \\
 isot &  2He4 & 1.31E-04 & 1.22E-04 & 1.00E-04 & 4.64E-05 & 3.05E-08 & 1.32E-04 & 1.23E-04 & 1.01E-04 & 4.62E-05 & 3.18E-08 \\
 elem &  6C   & 5.02E-03 & 4.94E-03 & 4.85E-03 & 5.16E-03 & 5.19E-03 & 2.40E-03 & 2.37E-03 & 2.55E-03 & 2.50E-03 & 2.40E-03 \\
 isot &  6C12 & 5.02E-03 & 4.94E-03 & 4.85E-03 & 5.16E-03 & 5.19E-03 & 2.40E-03 & 2.37E-03 & 2.55E-03 & 2.50E-03 & 2.40E-03 \\
 isot &  6C13 & 2.08E-10 & 3.58E-09 & 1.25E-08 & 3.02E-08 & 7.54E-08 & 1.06E-10 & 1.58E-09 & 6.48E-09 & 1.52E-08 & 3.90E-08 \\
\hline
\end{tabular}
\vspace{-0.5cm}
\footnotetext{Sample of Table~\ref{tab:nuc-chan-std}, the full version is available online. The meaning of 
the columns is the same as in Table~\ref{tab:nuc-chan}.}
\end{minipage}
\end{table*}

\begin{table*}
\begin{minipage}{210mm}
\caption{Nucleosynthesis in sub-Chandrasekhar models with standard $^{12}$C+$^{16}$O reaction rate.}
\label{tab:nuc-subch-std}
\begin{tabular}{llllllllllll}
\hline
 \multicolumn{2}{l}{$M_\mathrm{WD}$} &              0.88 &     0.88 &     0.88 &     0.88 &     0.88 &     
0.97 &     0.97 &     0.97 &     0.97 &     0.97 \\
 \multicolumn{2}{l}{$Z$} &              2.25E-4 &  2.25E-3 &  9.00E-3 &  2.25E-2 &  6.75E-2 &  
2.25E-4 &  2.25E-3 &  9.00E-3 &  2.25E-2 &  6.75E-2 \\
 elem &  2He  & 9.87E-08 & 9.71E-08 & 7.65E-08 & 3.33E-08 & 2.92E-10 & 2.25E-04 & 2.39E-04 & 2.72E-04 & 3.35E-04 & 4.93E-04 \\
 isot &  2He3 & 1.97E-16 & 1.97E-16 & 1.99E-16 & 2.04E-16 & 2.39E-16 & 7.16E-16 & 7.09E-16 & 6.89E-16 & 6.52E-16 & 5.39E-16 \\
 isot &  2He4 & 9.87E-08 & 9.71E-08 & 7.65E-08 & 3.33E-08 & 2.92E-10 & 2.25E-04 & 2.39E-04 & 2.72E-04 & 3.35E-04 & 4.93E-04 \\
 elem &  6C   & 4.10E-03 & 4.08E-03 & 3.99E-03 & 3.83E-03 & 3.33E-03 & 1.67E-03 & 1.66E-03 & 1.63E-03 & 1.56E-03 & 1.36E-03 \\
 isot &  6C12 & 4.10E-03 & 4.08E-03 & 3.99E-03 & 3.83E-03 & 3.33E-03 & 1.67E-03 & 1.66E-03 & 1.62E-03 & 1.55E-03 & 1.36E-03 \\
 isot &  6C13 & 2.25E-10 & 3.71E-09 & 1.23E-08 & 2.61E-08 & 5.62E-08 & 9.58E-11 & 1.53E-09 & 5.68E-09 & 1.28E-08 & 2.83E-08 \\
\hline
\end{tabular}
\vspace{-0.5cm}
\footnotetext{Sample of Table~\ref{tab:nuc-subch-std}, the full version is available online. The meaning of 
the columns is the same as in Table~\ref{tab:nuc-chan}.}
\end{minipage}
\end{table*}

\begin{table*}
\begin{minipage}{210mm}
\caption{Radioactivities with half-life longer than one day in Chandrasekhar-mass models with the standard $^{12}$C+$^{16}$O reaction rate.}
\label{tab:rad-chan-std}
\begin{tabular}{lllllllllll}
\hline
 \multicolumn{1}{l}{$\rho_\mathrm{DDT}$} &         1.2E+07 &  1.2E+07 &  1.2E+07 &  1.2E+07 &  1.2E+07 &  
1.6E+07 &  1.6E+07 &  1.6E+07 &  1.6E+07 &  1.6E+07 \\
 \multicolumn{1}{l}{$Z$} &              2.25E-4 &  2.25E-3 &  9.00E-3 &  2.25E-2 &  6.75E-2 &  2.25E-4 &  
2.25E-3 &  9.00E-3 &  2.25E-2 &  6.75E-2 \\
 Al26 &   1.29E-07 & 3.89E-07 & 5.25E-07 & 8.19E-07 & 1.32E-06 & 8.06E-08 & 2.01E-07 & 2.86E-07 & 3.97E-07 & 5.71E-07 \\
 P 32 &   2.25E-10 & 2.97E-08 & 1.57E-07 & 5.68E-07 & 4.76E-06 & 1.40E-10 & 2.00E-08 & 1.05E-07 & 3.41E-07 & 2.57E-06 \\
 P 33 &   8.13E-11 & 2.34E-08 & 1.49E-07 & 5.37E-07 & 3.73E-06 & 5.02E-11 & 1.38E-08 & 9.17E-08 & 3.13E-07 & 2.10E-06 \\
 S 35 &   2.74E-11 & 6.40E-09 & 7.57E-08 & 7.14E-07 & 1.08E-05 & 1.58E-11 & 4.10E-09 & 4.96E-08 & 4.22E-07 & 5.93E-06 \\
 Ar37 &   6.31E-06 & 1.08E-05 & 2.22E-05 & 3.74E-05 & 6.53E-05 & 4.11E-06 & 7.87E-06 & 1.65E-05 & 2.73E-05 & 4.77E-05 \\
 Ca41 &   6.71E-07 & 1.72E-06 & 4.13E-06 & 7.20E-06 & 1.08E-05 & 5.64E-07 & 1.35E-06 & 3.19E-06 & 5.38E-06 & 7.94E-06 \\
\hline
\end{tabular}
\vspace{-0.5cm}
\footnotetext{Sample of Table~\ref{tab:rad-chan-std}, the full version is available online. The meaning of the 
columns is the same as in Table~\ref{tab:rad-chan}.}
\end{minipage}
\end{table*}

\begin{table*}
\begin{minipage}{210mm}
\caption{Radioactivities with half-life longer than one day in sub-Chandrasekhar models with the standard $^{12}$C+$^{16}$O reaction rate.}
\label{tab:rad-subch-std}
\begin{tabular}{lllllllllll}
\hline
 \multicolumn{1}{l}{$M_\mathrm{WD}$} &              0.88 &     0.88 &     0.88 &     0.88 &     0.88 &     
0.97 &     0.97 &     0.97 &     0.97 &     0.97 \\
 \multicolumn{1}{l}{$Z$} &              2.25E-4 &  2.25E-3 &  9.00E-3 &  2.25E-2 &  6.75E-2 &  2.25E-4 &  
2.25E-3 &  9.00E-3 &  2.25E-2 &  6.75E-2 \\
 Al26 &   9.89E-08 & 2.81E-07 & 3.74E-07 & 5.25E-07 & 7.03E-07 & 5.46E-08 & 1.07E-07 & 1.40E-07 & 1.87E-07 & 2.30E-07 \\
 P 32 &   1.68E-10 & 2.37E-08 & 1.23E-07 & 4.09E-07 & 2.91E-06 & 8.68E-11 & 1.24E-08 & 5.96E-08 & 1.84E-07 & 1.17E-06 \\
 P 33 &   5.59E-11 & 1.87E-08 & 1.16E-07 & 3.90E-07 & 2.38E-06 & 2.93E-11 & 8.08E-09 & 4.81E-08 & 1.62E-07 & 9.77E-07 \\
 S 35 &   1.93E-11 & 5.13E-09 & 5.74E-08 & 4.85E-07 & 6.31E-06 & 8.90E-12 & 2.83E-09 & 2.96E-08 & 2.27E-07 & 2.68E-06 \\
 Ar37 &   4.82E-06 & 8.15E-06 & 1.69E-05 & 2.76E-05 & 4.56E-05 & 3.37E-06 & 6.63E-06 & 1.32E-05 & 2.16E-05 & 3.59E-05 \\
 Ca41 &   4.95E-07 & 1.31E-06 & 3.16E-06 & 5.30E-06 & 7.52E-06 & 5.09E-07 & 1.19E-06 & 2.62E-06 & 4.35E-06 & 6.19E-06 \\
\hline
\end{tabular}
\vspace{-0.5cm}
\footnotetext{Sample of Table~\ref{tab:rad-subch-std}, the full version is available online. The meaning of the 
columns is the same as in Table~\ref{tab:rad-chan}.}
\end{minipage}
\end{table*}

\section{Hydrodynamics and nucleosynthesis code}\label{appb}

The supernova code solves the hydrodynamics and nucleosynthesis along the explosion in a single run, i.e. 
with no postprocessing of the nuclear equations. This approach has the advantage that the binding energy 
released in the nuclear processes is treated self-consistently in the hydro part of the problem, as well as 
the dependence of the equation of state on the chemical composition. As an example, during alpha-rich 
freeze-out from nuclear statistical equilibrium (NSE) the specific nuclear binding energy can change as much 
as $9\times10^{16}$~erg~g$^{-1}$ ($\sim10\%$), while the mean molar weight varies from $\sim30 - 32$ at the 
beginning of the freeze-out up to $\sim52 - 54$ at the end of the same period.

The hydrodynamic solver is explicit and based on the piecewise-parabolich method (PPM) in the Lagrangian 
version of \cite{1984col}, with spherical symmetry and gravity. We adopt the modifications introduced 
in \cite{1985col} to solve the Riemann problem with a general equation of state (EOS).
The integration of the hydrodynamic equations with the PPM scheme is followed by the integration of the nuclear kinetic equations, and then the temperature is updated according to the nuclear energy released (and to the energy loss rate due to neutrinos generated in weak transitions).
We take profit of the 
flattening algorithms introduced by \cite{1984col}, in order to identify the mass shells affected by shocks during 
the detonation phase of the SN Ia explosion, and we forbid nuclear burning until the shock has passed away 
from the shell. This procedure mimicks the expected behaviour of actual detonations, in which the dissipation associated with the 
leading shock front heats matter almost instantaneously to temperatures on the order of $3 - 4\times10^9$~K 
(depending on the fuel density), with no change in the chemical composition. 

Our EOS accounts for radiation and for fully ionized matter: electrons and positrons \citep{1996bli}, 
relativistic and non-relativistic with arbitrary degree of degeneracy, and
ions, with Coulomb and polarization corrections  
from \citet{1989yak} and \citet{1987oga}, see also \citet{1999bra}.
The information about the chemical composition is fully shared between the nuclear solver and the hydro solver. It has an imprint, for instance, in the Coulomb corrections to the ionic EOS.

Within a thermonuclear supernova explosion, matter changes from a state of extreme degeneracy to partial and weak
degeneracy. We integrate a temperature, $T$, equation whenever matter is strongly degenerate, and a specific energy, $e$, 
equation otherwise. In practice, the first option is chosen when
\begin{equation}
 \left| \frac{\left(\partial e /\partial\log T\right) }{ \left(\partial e /\partial\log\rho\right) }
\right| < 0.01\,.
\end{equation}

\subsection{Spacing and time steps}

\begin{figure}\label{figb1}
   \includegraphics[width=\columnwidth]{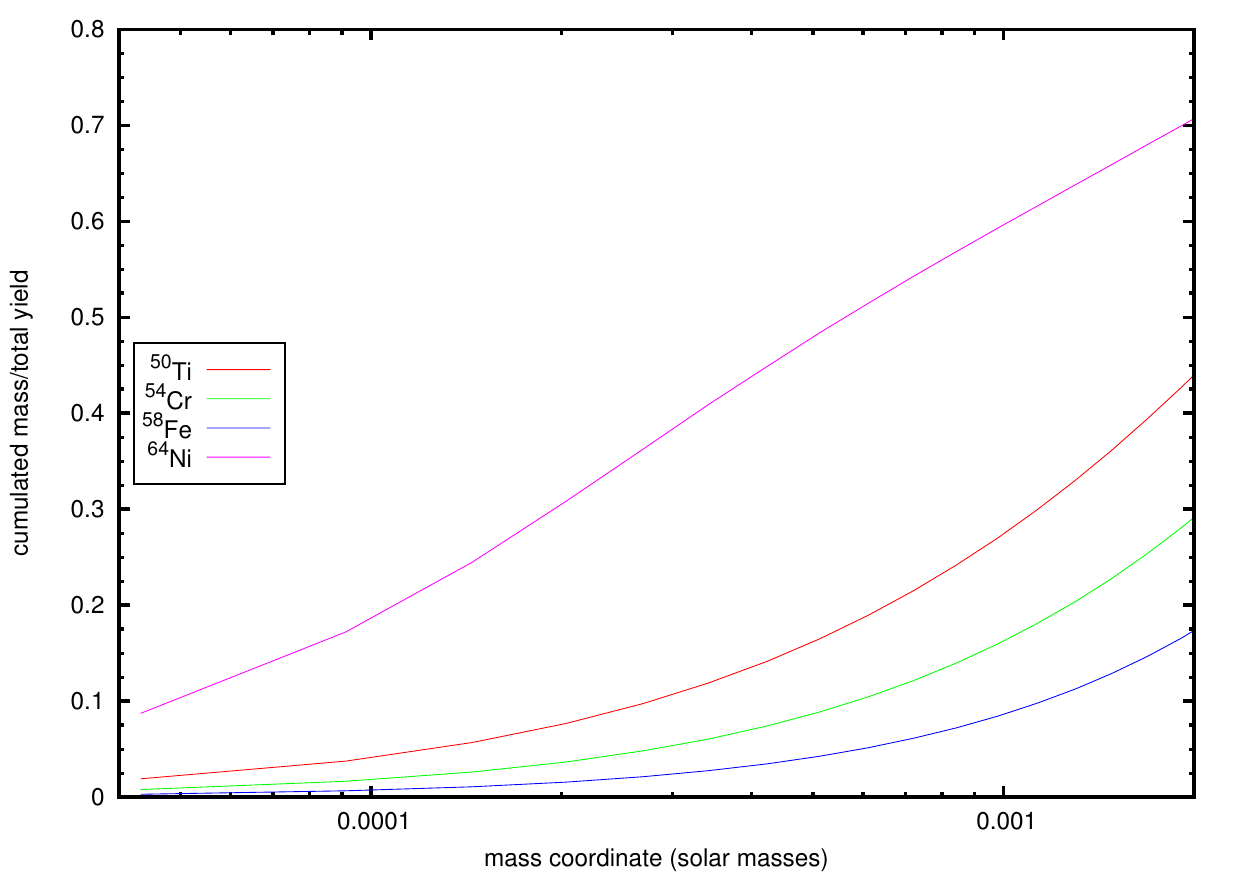}
    \caption{Final abundance profile of several neutronized isotopes close to the center of the WD in model 
ddt2p4\_Z9e-3\_$\xi_\mathrm{CO}$0p9. The ordinate gives the integrated mass, starting from the center, 
normalized by the total yield of the isotope. It shows that the adopted spatial resolution is fine enough to 
smoothly represent the production of \isotope{50}{Ti}, \isotope{54}{Cr}, and \isotope{58}{Fe}. The central 
layer contributes slightly less than $10\%$ to the total mass of \isotope{64}{Ni}, which is anyway 
underproduced in this model by a factor $\gtrsim2$ (see, e.g., Fig.~\ref{fig12a}).}
\end{figure}

The shell spacing, in terms of Lagrangian mass coordinate, is uniform and equal to $0.002\times\mwd$ in all the star but for the innermost and outermost $0.02\times\mwd$, 
where it is finer. Starting from a central mass shell of $3\times10^{-5}\times\mwd$, the mass of each 
shell increases by 10\% until reaching the width of $0.002\times\mwd$. 
In the outermost layers, we apply the 
same procedure as in the center but in reverse order. In total, there are 562 mass shells in each model. Figure~\ref{figb1} shows the final abundance profile of the isotopes that 
are most contributed by the central layers in a delayed-detonation model, where it can be seen that their 
production is well resolved by the adopted spatial zoning. The fine zoning applied in the outer layers 
contributes to reproducing acurately the density gradient and the curvature effect on the propagation of the 
detonation front \citep{2001sha,2013dun,2018mil}. As an example, at the time of formation of the detonation 
in model ddt2p4\_Z9e-3\_std, the spatial resolution is better than $\Delta R/R < 1\%$ in all the WD, and the 
relative change in density between adjacent fuel layers is less than $\Delta\rho/\rho < 10\%$ in all but the 
outermost $10^{-4}~\msun$.

The time step is adaptive to ensure that the integration variables do not change too much in a single time 
step. The variables we control are density, temperature, velocity, abundances, and shell sound crossing 
time (Courant condition). The time step in the hydrodynamic solver, $\Delta t$, can be different from that of the nuclear network solver, $\Delta t_\mathrm{nuc}$. Besides, the value of $\Delta t$ is common to all mass shells at any given integration step, while the value of $\Delta t_\mathrm{nuc}$ can differ from shell to shell. Specifically, we require for each hydrodynamic time step, that the relative change 
is smaller than $0.1\%$ in density, $0.2\%$ in temperature, and $20\%$ in velocity (with a floor of 
$10^6$~\cms). With respect to the nuclear network time step, we accept a relative change up to $2\%$ in the abundance of any nuclear species whose molar abundance is larger than 
$10^{-6}$~mol~g$^{-1}$ in any mass shell. If the new time step results smaller than $40\%$ of the previous 
one, the integration of the last time step, either a hydrodynamic or a nuclear one, is rejected and repeated with a smaller time step, until the above 
condition is fulfilled. The repetition of a hydrodynamic integration step implies that of the nuclear solver for all mass shells. On the other hand, the repetition of a nuclear integration step does not require that of the hydrodynamic solution. 

When the nuclear time step is smaller than the current hydrodynamic time step, the nuclear evolution is solved with frozen temperature and density until the accumulated nuclear time equals the hydrodynamic time step. After accounting for the nuclear energy released, sometimes the condition of maximum relative change for the temperature is not fulfilled, in which case the whole process is repeated starting from the last hydro and nuclear good values, but with a smaller $\Delta t$. 
In Fig.~B2, 
there can be seen an example of the run of the hydrodynamic time step along 
the explosion.

\begin{figure}\label{figb2}
   \includegraphics[width=\columnwidth]{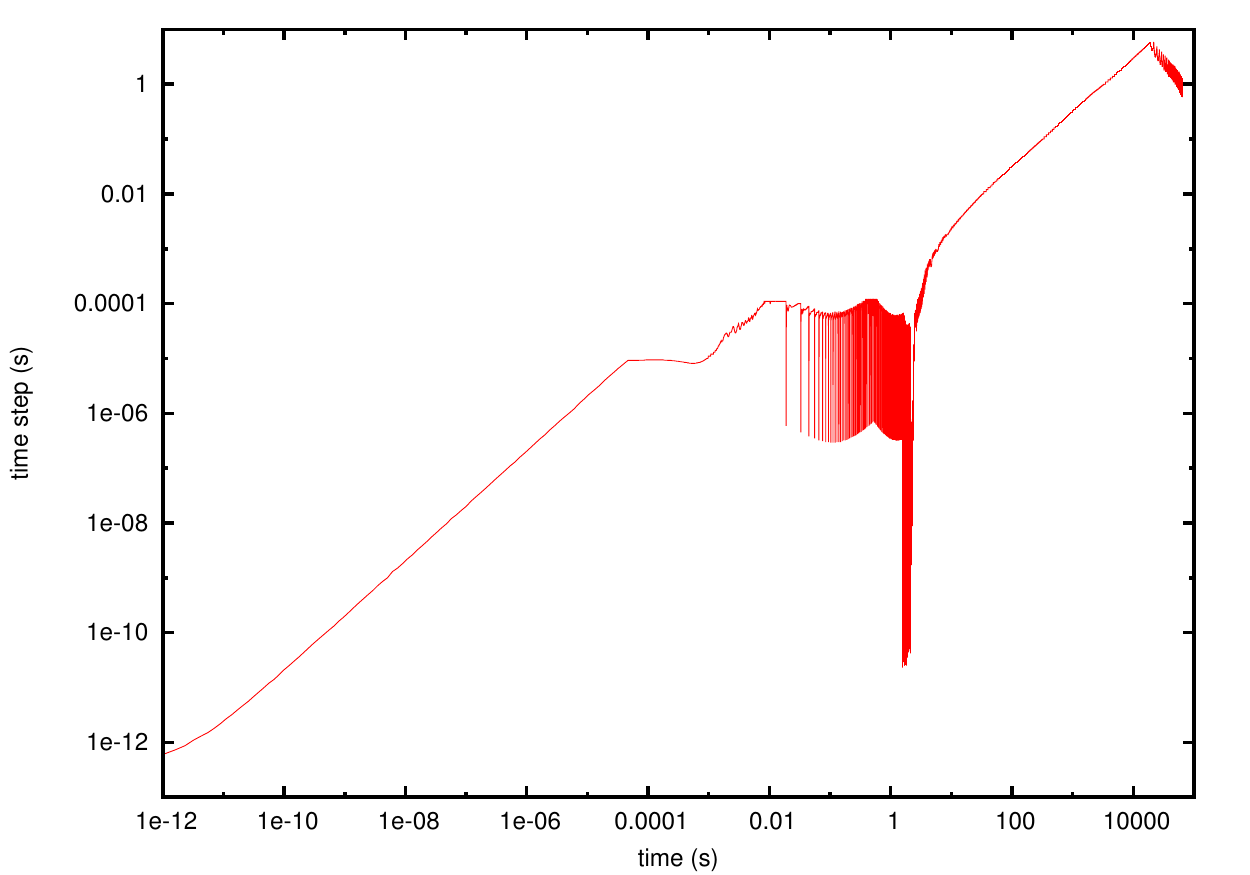}
    \caption{Hydrodynamic time step along the explosion of model ddt2p4\_Z9e-3\_std. The sudden decreases on 
the time step down to $\Delta t\sim10^{-7}$~s from time $\sim0.02$~s to $\sim1.5$~s belong to the arrival of 
the deflagration front to a new mass shell when the fuel density is high enough to assume a NSE state will 
be reached. In our models, the minimum fuel density at which NSE is forced is $8\times10^7$~\gcc. Thereafter, 
the more pronounced decreases on the time step down to $\Delta t\sim10^{-11}$~s reflect the arrival of  
either the flame front or the detonation wave (through shock heating) to a new mass shell, and the subsequent 
acceleration of the nuclear reactions.}
\end{figure}

\subsection{Nuclear network}

The nuclear network, shown in Fig.~B3, 
is the same as in \cite{2012bra}. The network extends to sufficiently neutron-rich isotopes, beyond the valley of beta stability, to describe the high-density combustion phases, in which electron captures increase the mean neutron excess of matter to $\sim0.1$ (electron mole number, $Y_\mathrm{e}\sim0.44$).
The nuclear kinetic equations follow the time evolution of the mass fraction, $X_i$, 
of each species due to weak and strong interactions, including photodisintegrations and four fusion reactions
($3\alpha$, $^{12}$C+$^{12}$C, $^{12}$C+$^{16}$O, and $^{16}$O+$^{16}$O), until the temperature
falls below $10^8$~K, after which time only the weak interactions are accounted for.


\begin{figure}\label{figb3a}
   \includegraphics[width=\columnwidth]{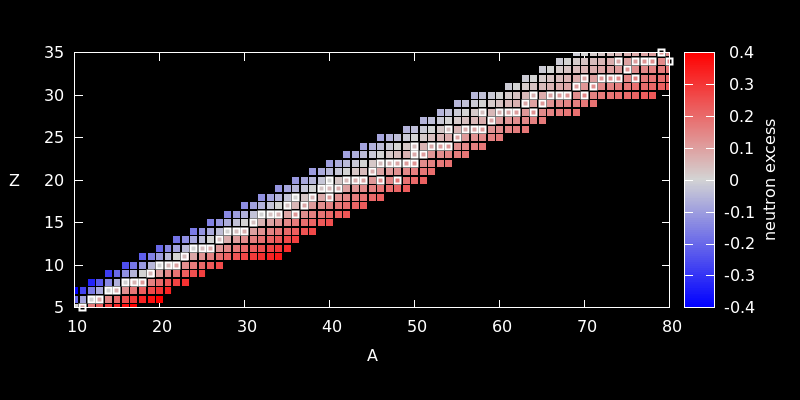}
    \caption{Nuclear network used in the hydrodynamic calculations reported in this paper.
Location of the main contributors to nucleosynthesis, besides p, n, and $\alpha$, on the charge -- baryon number plane, 
coloured according to their neutron
excess (the actual network is slightly larger, from $Z=0$ to $Z=50$ and from $A=1$ to $A=100$). We highlighted the stable 
isotopes with an open white square.
}
\end{figure}

\begin{figure}\label{figb3b}
   \includegraphics[width=\columnwidth]{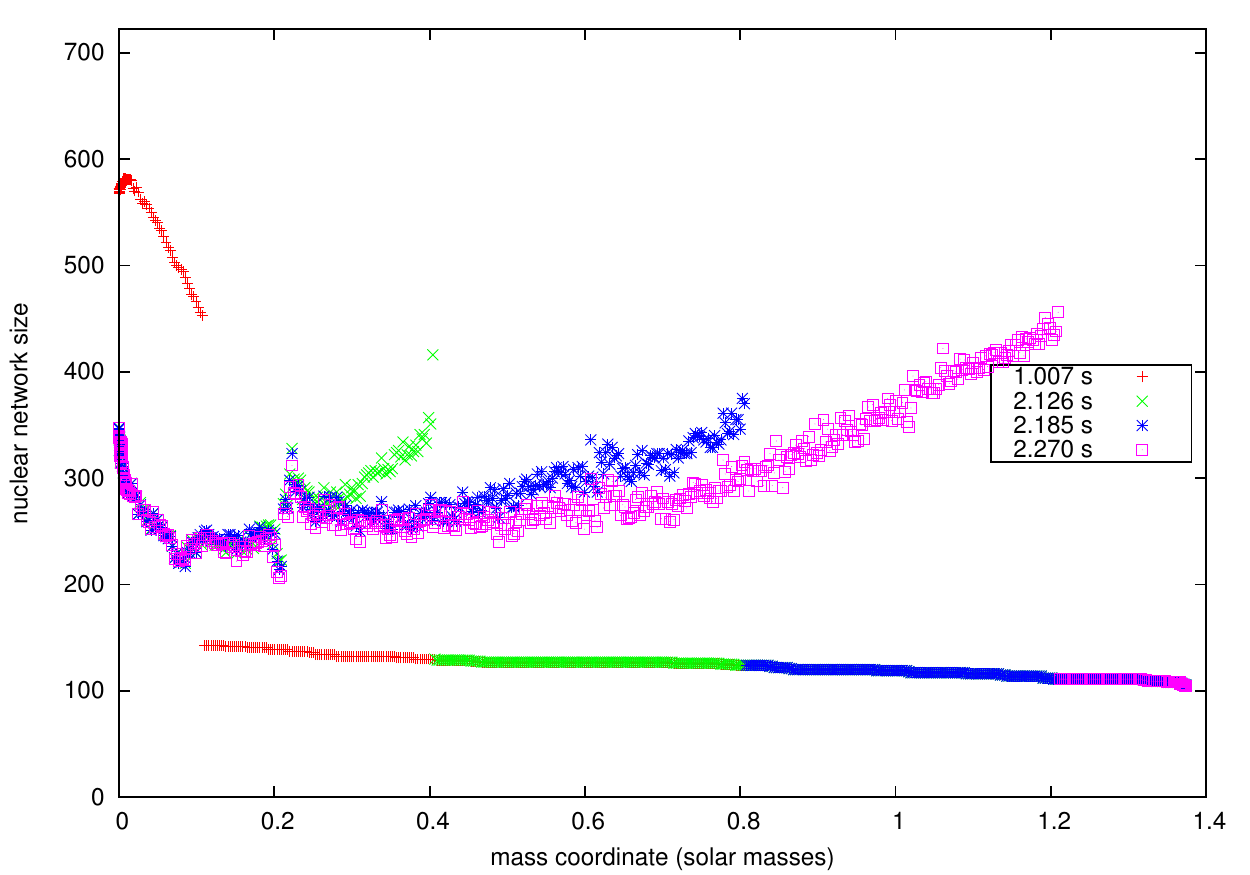}
    \caption{
Number of nuclei actually present in the computational network as function of the Lagrangian mass coordinate in model ddt2p4\_Z9e-3\_std at four different times. 
At 1.007~s the explosion is still in the deflagration phase and registers the largest network size close to the center, where matter is in the NSE state. 
The other three times depicted, 2.126~s, 2.185~s and 2.270~s, belong to the detonation phase and show increasing network size close to the detonation front, and close to the 
WD surface as well. In each shell, the network size remains constant, on the order of 100 -- 150 species, until the flame front reaches it, which results in overlapping of the different symbols at the plot bottom.
}
\end{figure}

\begin{figure}\label{figb3c}
   \includegraphics[width=\columnwidth]{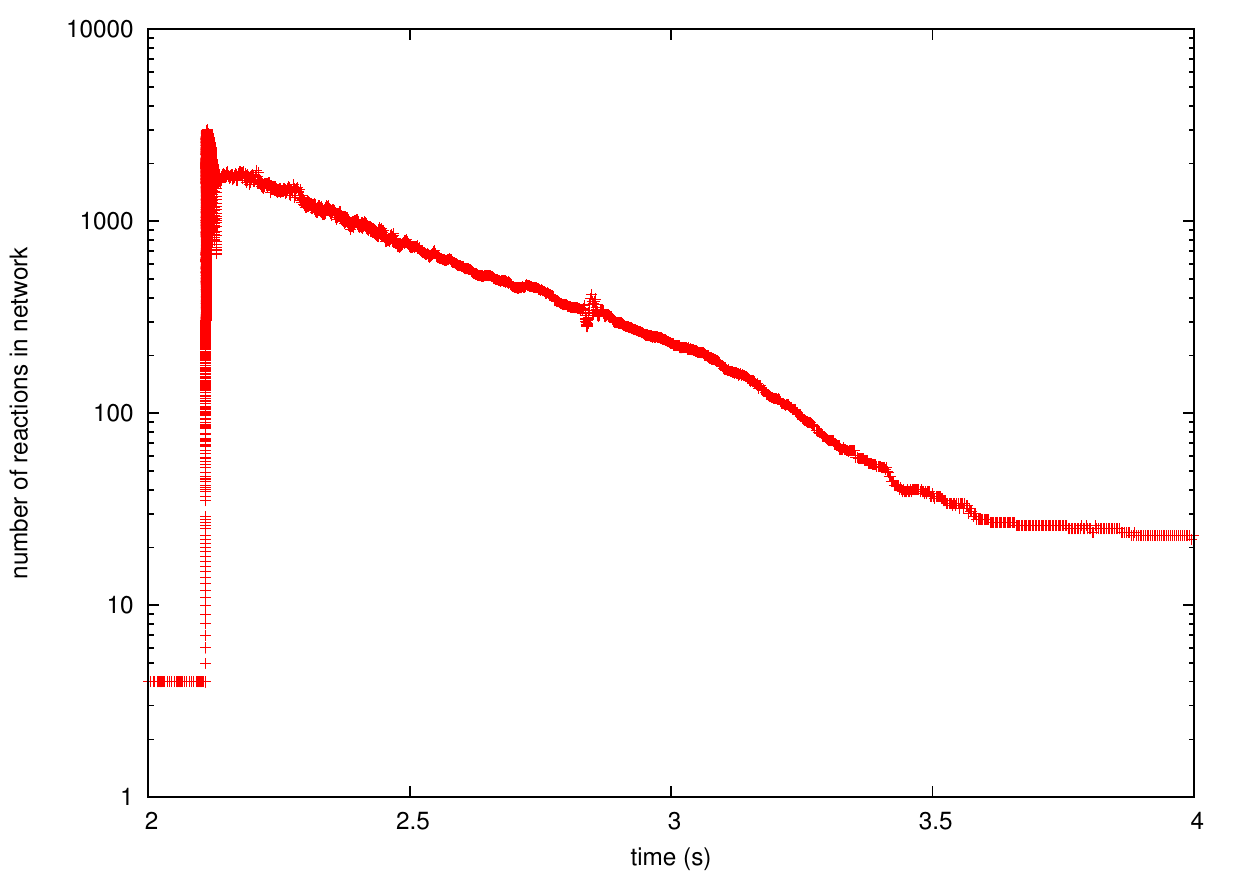}
    \caption{
Time evolution of the number of nuclear reactions (strong and weak, including photodisintegrations and fusion reactions)
in the network for the mass shell with Lagrangian mass coordinate 0.3~\msunb  in model ddt2p4\_Z9e-3\_std.
}
\end{figure}

The nuclear network solver is implicit, iterative, and uses adaptive time steps.
In each time step, the iterative procedure ends when the molar abundances of all nuclei
with $Y_i>10^{-17}$~mol~g$^{-1}$ have converged to better than  
$10^{-4}\times\mathrm{max}(Y_i,10^{-6}~\mathrm{mol}~\mathrm{g}^{-1})$.

The size and composition of the nuclear network changes each time step and may be different from mass shell to 
mass shell (see Fig.~B4), 
with a maximum of 722 nuclides. 
The nuclides actually present in the network are determined according to the chemical abundances and the 
possible nuclear links to other abundant nuclei. Initially, the network is defined by neutrons and the rest of isotopes present in the 
Solar System mixture up to  $^{101}$In. Thereafter, the network is formed by neutrons, protons, alphas, and those species with
abundance $Y_i > Y_\mathrm{thresh} = 10^{-19}$~mol~g$^{-1}$, plus the nuclei that can be reached from any of the
nuclei with $Y_i > 100\times Y_\mathrm{thresh}$  by any one of the reactions included in the network.
A reaction rate is included in the network only if the predicted change of a molar abundance in the next time step, $\Delta t_\mathrm{nuc}$, is
larger than a given threshold:
\begin{equation}
  N_{\mathrm{A}}\rho\langle\sigma v\rangle Y_i Y_j \Delta t_\mathrm{nuc} > R_\mathrm{thresh} = 10^{-13}~\mathrm{mol}~\mathrm{g}^{-1}\,.
\label{eqthreshold}
\end{equation}
\noindent We show in Fig.~B5 
how the number of reactions in the network changes 
along the explosion for a typical mass shell. 
A similar method of integration of the nuclear evolutionary equations using an adaptive network 
has been described in \cite{2002rau} and \cite{2012bra}, the main
novelty here is that it is solved along with the hydrodynamic evolution of the exploding WD. 

\begin{table*}
\caption{Convergence of nucleosynthesis calculations. The figures belong to model model ddt2p4\_Z9e-3\_std.}
 \label{tab:conver}
\renewcommand{\thefootnote}{\alph{footnote}}
 \begin{tabular}{@{}lllll@{}}
 \hline\hline
 & range of yield masses & 33\% of nuclides & 67\% of nuclides & 100\% of nuclides \\
 & (\msun) & better than & better than & better than \\
 \hline
$Y_\mathrm{thresh}=10^{-20}$~mol~g$^{-1}$ & $M_i \ge 10^{-3}$ & 0.001\% & 0.001\% & 0.07\% \\
{\hfill vs.\hfill} & $10^{-3}>M_i\ge10^{-6}$ & 0.001\% & 0.06\% & 2.7\% \\
$Y_\mathrm{thresh}=10^{-19}$~mol~g$^{-1}$  & $10^{-6}>M_i\ge10^{-12}$ & 0.001\% & 0.04\% & 2.2\% \\
\hline
$R_\mathrm{thresh} = 10^{-14}~\mathrm{mol}~\mathrm{g}^{-1}$ & $M_i \ge 10^{-3}$ & 0.001\% & 0.02\% & 0.25\% \\
{\hfill vs.\hfill} & $10^{-3}>M_i\ge10^{-6}$ & 0.06\% & 0.15\% & 3.3\% \\
$R_\mathrm{thresh} = 10^{-13}~\mathrm{mol}~\mathrm{g}^{-1}$  & $10^{-6}>M_i\ge10^{-12}$ & 0.6\% & 3.6\% & 28\% \\
\hline
weak rates table in S09 & $M_i \ge 10^{-3}$  & 0.4\% & 2.5\% & 65\% \\
{\hfill vs.\hfill} & $10^{-3}>M_i\ge10^{-6}$ & 0.5\% & 1.9\% & 480\% \\
our standard procedure & $10^{-6}>M_i\ge10^{-12}$ & 0.4\% & 2.2\% & 5020\% \\
 \hline\hline
 \end{tabular}
\end{table*}

\subsection{Convergence of the nucleosynthesis calculation}

In Table~\ref{tab:conver}, we show the convergence of the nucleosynthesis results for model ddt2p4\_Z9e-3\_std with respect to the chosen threshold values, 
$Y_\mathrm{thresh}$ and $R_\mathrm{thresh}$,
when any one of them is a factor ten smaller than our reference values, $Y_\mathrm{thresh} = 10^{-19}$~mol~g$^{-1}$ and $R_\mathrm{thresh} = 10^{-13}~\mathrm{mol}~\mathrm{g}^{-1}$, while all other parameters are held constant. 
For either an abundance threshold or a reaction rate 
threshold an order of magnitude smaller than our reference values, 
the final kinetic energy changes by less than 0.09\%, and the mass of \isotope{56}{Ni} synthesized by less than 0.02\%. Concerning the
yields of the different nuclides, with respect to the modification of $Y_\mathrm{thresh}$ by a factor ten, the yields change by at most 
0.001\% for 33\% of all nuclei and for 67\% of nuclei with a mass yield, $M_i$, larger than $10^{-3}$~\msun, while no nuclei with 
$M_ i > 10^{-3}$~\msunb changes by more than 0.07\%, no nuclei with $M_ i > 10^{-6}$~\msunb changes by more than 2.7\% 
(\isotope{61}{Ni} is the one with the largest variation), and no nuclei with $10^{-6}~\msunb > M_ i > 10^{-12}$~\msunb changes by more 
than 2.2\% (\isotope{47}{Ti}). The records are similar although slightly worse when we consider the modification of $R_\mathrm{thresh}$ by a
factor of ten. In this case, the largest variation belongs to \isotope{47}{Ti}, whose mass yield is $\sim10^{-7}$~\msun.

\subsection{Treatment of matter in NSE}

At temperatures in excess of $\sim5-6\times10^9$~K and suficiently high densities, most direct and reverse strong interactions achieve equilibrium, the NSE state, and the 
composition is determined by a Saha equation accounting for the nuclear mass, the nuclear partition function, density, temperature, and electron mole number. 
In our code, we assume matter achieves NSE when the burning front arrives to a mass shell with density $\rho \ge  \rho_\mathrm{NSE0} =  8\times10^7$~\gcc. Below this density,
we require a minimum temperature of $T \ge T_\mathrm{NSE1} = 5.5\times10^9$~K if density is $\rho \ge \rho_\mathrm{NSE1} = 2\times10^7$~\gcc, or  $T \ge T_\mathrm{NSE2} = 6\times10^9$~K otherwise, 
to assume an NSE state is achieved. Shells processed by the detonation front are not assumed to achieve NSE anyway 
if their density is smaller than  $\rho_\mathrm{NSE0}$, 
irrespective of how high is their temperature.
Instead their 
chemical evolution is followed with the nuclear network all the way. In practice, detonated shells are 
treated through the NSE routine only in a very limited set of models, because either the 
deflagration-to-detonation transition density of Chandra models stays below $\rho_\mathrm{NSE0}$, or the 
whole subCh initial models stay below the same density, even at the center of the WD (see Table~\ref{tab1}). 

Shells that have been incinerated to NSE are assumed to stay in the statistical equilibrium state (and their 
chemical composition in NSE is recalculated at each time step with the new $\rho$, $T$, and $Y_\mathrm{e}$) 
when their temperature is larger than
$T \ge T_\mathrm{out1} = 5\times10^9$~K if density is $\rho \ge \rho_\mathrm{out1} = 5\times10^6$~\gcc, or  $T \ge T_\mathrm{out2} = 5.8\times10^9$~K if density is 
$\rho \ge \rho_\mathrm{out2} = 10^6$~\gcc. Otherwise, i.e. for smaller temperatures, their abundances are fed to the nuclear network and their chemical evolution is followed solving the nuclear kinetic equations.

In thermonuclear supernova explosions, matter is neutronized by electron captures in the NSE state at high densities. In our code, the electron captures in NSE are calculated at each time step
according to the NSE composition and the weak interaction rates on each nuclei. The same method is applied to obtain the neutrino cooling due to weak interactions in NSE. 

Since we do not rely on 
tabulated neutronization rates, our code is well suited to rate the precision of the available tables of electron captures in NSE or, more precisely, how they impact on the results of the explosion. As an 
example, we have recomputed the hydrodynamics and nucleosynthesis of model ddt2p4\_Z9e-3\_std by using the 
extensive NSE tables provided by \citet[][hereafter S09]{2009se2}, and show the results in the last row of 
Table~\ref{tab:conver}. 
The kinetic energy and the yield of \isotope{56}{Ni} we obtain match nicely those calculated without the 
weak table, i.e. following our standard procedure of computing the NSE composition, weak rates, and neutrino emission coupled to the hydrodynamic solver. The discrepancy in kinetic energy is of order 0.07\%, comparable to our precision with respect 
to the parameters $Y_\mathrm{thresh}$ and $R_\mathrm{thresh}$, while the ejected mass of \isotope{56}{Ni} is 
only 0.7\% smaller than our standard value. 
On the other hand, the nucleosynthesis changes drastically with respect to our standard model. Within the most 
abundant nuclides, which we define as those with ejected mass $M_i > 10^{-3}\msunb$ for the present purposes, we find discrepancies 
up to $65\%$, among them the important SN Ia product \isotope{54}{Cr}. The yields of 
other neutronized nuclides such as \isotope{58}{Fe} and \isotope{62}{Ni} disagree from our standard model 
also by more than $10\%$. In the second row, that of nuclides whose yield is in the range 
$10^{-3}\msunb\ge M_i > 10^{-6}\msunb$, the largest discrepancy belongs to \isotope{64}{Ni}, whose yield 
differs from our standard model by as much as $480\%$, while the discrepancy of the yield of \isotope{50}{Ti} 
is also above $100\%$. Finally, within the nuclides with $M_i > 10^{-12}\msunb$, the largest discrepancy 
amounts to $5020\%$, which is the case for \isotope{48}{Ca}.
We have checked that our neutronization rate and neutrino emission rate match those given in S09 
at the tabulated values of $T$, $\rho$, and $Y_\mathrm{e}$, so the different nucleosynthetic results we obtain are 
attributable uniquely to the interpolation between the nodes of their NSE tables.

\bsp	
\label{lastpage}
\end{document}